\documentclass[12pt,preprint]{aastex}
\newcommand{\grad}{\mbox{\boldmath$\nabla$}}
\renewcommand{\v}{{\mathbf{v}}}
\newcommand{\vdot}{{\mathbf{\cdot}}}
\newcommand{\vcross}{{\mathbf{\times}}}
\newcommand{\thth}{\hspace{1.5pt}}
\newcommand\Div{\grad\vdot\thth}

\shortauthors{Dikpati &  Gilman}
\shorttitle{High latitude meridional circulation}
\begin{document}

\title{THEORY OF SOLAR MERIDIONAL CIRCULATION AT HIGH LATITUDES}

\author{Mausumi Dikpati and Peter A. Gilman }

\affil{High Altitude Observatory, National Center for Atmospheric
Research, 3080 Center Green, Boulder, CO 80307-3000.}

\email{dikpati@ucar.edu,gilman@ucar.edu}

\begin{abstract}

We build a hydrodynamical model for computing and understanding the Sun's
large-scale high latitude flows, including Coriolis forces, turbulent
diffusion of momentum and gyroscopic pumping. Side boundaries of
the spherical 'polar cap', our computational domain, are located
at latitudes $\geq 60^{\circ}$. Implementing observed low latitude
flows as side boundary conditions, we solve the flow equations for a
cartesian analog of the polar cap. The key parameter that
determines whether there are nodes in the high latitude meridional
flow is $\epsilon=2 \Omega n \pi H^2/\nu$, in which $\Omega$ is
the interior rotation rate, n the radial wavenumber of the
meridional flow, $H$ the depth of the convection zone and $\nu$
the turbulent viscosity. The smaller the $\epsilon$ (larger
turbulent viscosity), the fewer the number of nodes in high
latitudes. For all latitudes within the polar cap, we find three
nodes for $\nu=10^{12}{\rm cm}^2{\rm s}^{-1}$, two for $10^{13}$, 
and one or none for $10^{15}$ or higher. For $\nu$ near $10^{14}$ 
our model exhibits 'node merging': as the meridional flow speed is 
increased, two nodes cancel each other, leaving no nodes. On the 
other hand, for fixed flow speed at the boundary, as $\nu$ is 
increased the poleward most node migrates to the pole and disappears, 
ultimately for high enough $\nu$ leaving no nodes. These results 
suggest that primary poleward surface meridional flow can extend from 
$60^{\circ}$ to the pole either by node-merging or by node migration 
and disappearance. 
\end{abstract}



\section{INTRODUCTION}
It has been demonstrated both observationally and theoretically that
meridional circulation plays a crucial role in the workings of the solar
cycle. In flux-transport dynamos, the meridional circulation is
responsible for determining the cycle period \citep{ws91,dc99,krs01}
and also plays an important role in setting the cycle 'shape', that is, 
its rise and fall patterns. It follows that it is very important for 
understanding and predicting solar cycles that we have the best 
possible information about meridional circulation on the Sun. The 
accurate measurements of surface flow-patterns as a function of latitude 
as well as time would be very useful for simulating the evolutionary 
pattern of the Sun's large-scale fields, particularly the polar fields 
\citep{bsss04,wls05,ddg08}.

\begin{figure}[hbt]
\epsscale{1.0}
\plotone{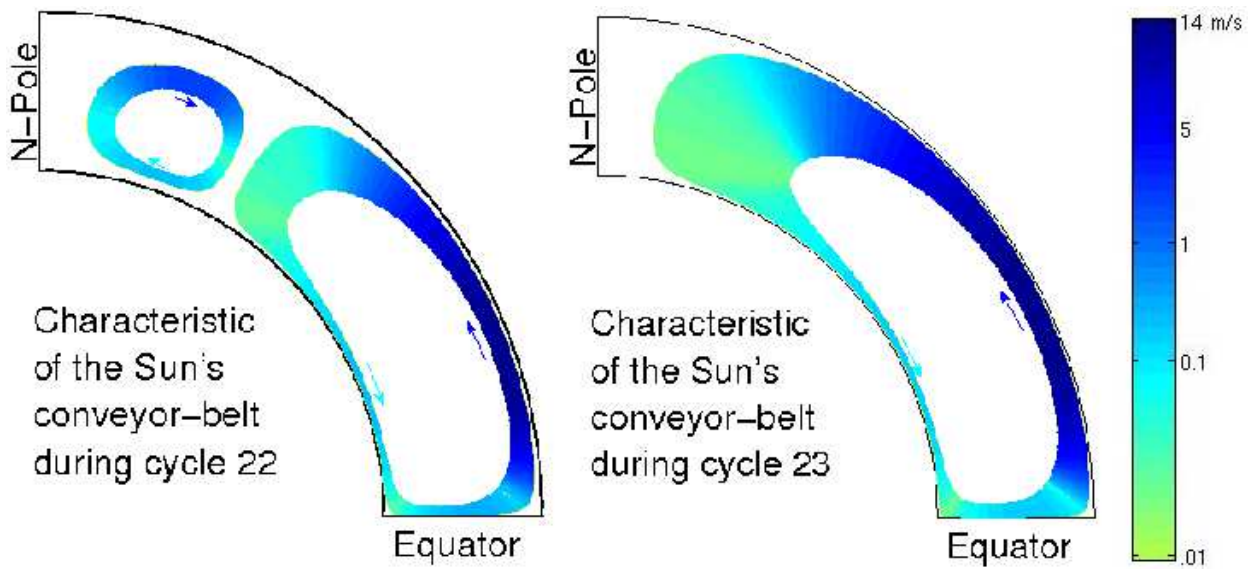}
\caption{The Sun's characteristic conveyor-belts during
cycles 22 (left frame) and 23 (middle frame), derived from surface
observations and mass-conservation, are shown. The speed associated
with the colors is shown in the colorbar (right frame). The maximum
surface flow-speeds at $25^{\circ}$ latitude were almost the same in both
the conveyor-belts, but the flow turning down towards the equator around
$60^{\circ}$ latitude during cycle 22 made the length of the primary
conveyor-belt shorter than that during cycle 23, in which flow went all
the way to the pole before turning equatorward.}
\label{mc2223}
\end{figure}

While meridional flow at all latitudes is important for the solar
dynamo, its pattern near the poles of the Sun may be particularly
significant. Recently \citet{dgdu10} showed the variation in
solar cycle duration could be explained by variations in the latitudinal
extent of the Sun's primary 'conveyor belt' or meridional circulation.
Observations of the Sun's surface Doppler plasma flow from Mount Wilson
Observatory data indicate that in both cycles 22 and 23, the maximum
poleward surface flow-speed in the primary belt was the same (\citet{ub05}; 
also see the figure 1 of \citet{dgdu10}). But in cycle 22, the primary 
circulation cell flowed poleward only to about $60^{\circ}$ latitude, 
thus making a shorter path for the magnetic flux transport via the 
conveyor belt and resulting in a cycle duration of $\sim 10.5$ years 
(see the left frame of Figure 1). On the other hand, in cycle 23, the 
poleward surface flow went all the way to the poles (see the middle frame 
in Figure 1), leading to a longer path via the conveyor-belt cycle of 
$\sim 12.5$ years.

In the past, some flux-transport dynamo calculations \citep{bebr05} 
and surface-transport calculations \citep{jcss08} have dealt with 
possible multi-cell meridional flow scenarios, in the context of understanding 
the role of these flows in solar cycle features. We now see that these 
studies are not just the merely playing with models; such scenarios 
could happen in reality. The change in the surface poleward flow-pattern -- 
its reversing around $60^{\circ}$ as it did in cycle 22 and maintaining 
poleward flow all the way to the pole as in cycle 23 -- have significantly 
impacted the duration of cycle 23 and the length of the minimum that followed 
it, compared to that in cycle 22.

The plasma velocity can also be determined by helioseismic analysis,
either from ring diagrams or from time-distance diagrams \citep{gdsb97,
hetal02,zk04,gkhhk08,gbs10}. Alternatively, features seen on the images 
such as magnetic structures and supergranule cells, can be tracked with  
cross-correlation analysis to yield a drift velocity for that feature 
\citep{khh93,sd96,sks06,szk07,sksb08}. \citet{u10} reanalyzed the Mount 
Wilson surface Doppler data for cycles 22 and 23, from 1986 through 
2009. \citet{u10} computed meridional flow profiles up to at least 
$80^{\circ}$ latitude, found smooth evolution of the signal from one 
year to the next, and confirmed the difference in high latitude flow 
patterns between cycles 22 and 23, in particular, the existence of a 
second reversed cell poleward of about $60^{\circ}$ during most of 
cycle 22, and a single cell with poleward surface flow all the way to 
the poles during the major part of cycle 23.  

Differential rotation throughout the solar convection zone is fairly 
well known from helioseismic measurements \citep{t04}. It is
difficult to measure at high latitudes because of foreshortening and 
other effects. Most methods measure the linear rotational velocity rather 
than the angular rotation rate, so it is particularly difficult to 
calculate the angular measure with the short moment-arm near the poles. 
It appears that on average the angular measure of differential rotation 
at the surface declines monotonically to the poles \citep{b00}, though 
the existence of a 'polar vortex' has been suggested by theory \citep{g79}. 
Although our focus on this paper is more on meridional flow at high latitudes 
than on differential rotation there, our model will calculate both quantities,
so we will need to compare results from the model with both flows.

Given the important consequences of a second, reversed meridional cell
in high latitudes, it is important to develop hydrodynamical theories that
could indicate what to expect to occur in the Sun. What physics determines the
presence or absense of a second cell? Can we develop a simple physical argument
that we should expect to see a single primary cell, or two or more cells? 
This is the question we attempt to answer in this paper. The theory we develop 
filters out all convective instability and concentrates on forcing high 
latitude meridional circulation mechanically by the primary meridional flow, 
and possibly by differential rotation, from lower latitudes. We recognize 
that in the Sun there could be both this mechanical forcing and axisymmetric 
convection in high latitudes.

Meridional circulation is produced in virtually all fluid dynamical models 
used to simulate and understand the origins of the differential
rotation of the Sun. These fluid dynamical models fall generally into
two classes: mean field models that are axisymmetric and include
parameterizations of turbulent transport of momentum \citep{drsc89,r05}, and
global 3D numerical models that simulate global
convection in a deep rotating spherical shell \citep{mbdt08}. 
For both types of models there are numerous results that show a
wide variety of meridional flow patterns, but recently both approaches
have been converging toward a common result of a dominant meridional
flow cell that has poleward flow near the outer boundary, and
equatorward return flow near the bottom \citep{r05,mbdt08}. 

For some parameter choices the mean field models give a second, reversed
cell at high latitudes \citep{r05} and the 3D global convection
models can give multiple high latitude cells \citep{mbdt08}. Both 
classes of models also show that the percentage fluctuations in the 
amplitudes of meridional flow with time are much larger than for the 
differential rotation. This appears to be due to time-fluctuations in the 
Reynolds stresses and the fact that the meridional flow is a result of a slight
imbalance between large forces, while the forces responsible for differential 
rotation are small enough to make the differential rotation change only 
very slowly.

These models are all global, and focused primarily on the problem of 
understanding the details of differential rotation with depth and
latitude in the Sun, as well as such features as torsional oscillations.
To focus primarily on the fluid dynamics of high latitudes, it is
possible to build simpler models, particularly at first, and then build
up to more realistic models more comparable to the global models
just described. This is the approach we take below.

\section{DEVELOPMENT OF THEORY}

As far as is known, all main sequence stars with outer convection zones rotate,
so they all have well defined equatorial planes and rotation poles.
They are also almost certain to have a meridional circulation. But depending on
the rotation rate, as well as the convection zone thickness, this circulation 
could differ greatly \citep{kr05,kr08}. For example, fast-rotating solar type 
stars may have meridional circulation whose primary cell has flow toward 
the equator rather than the poles as on the Sun \citep{rk02}. We know little
about meridional flow from observations in stars other than the Sun, but a 
comparision to planetary circulations makes the point. 

Most planetary atmospheres for which there are velocity measurements display
meridional circulation cells, the number of which in each hemisphere varies
significantly from planet to planet (due in part to the differences in 
planetary rotation rate) and with seasons as well as the presence or absence
of internal heat sources. Jupiter is a rather fast rotator, and it has many 
axisymmetric meridional cells between equator and pole (see \citet{kfs09}
and references therein). By contrast, Venus, a slow rotator, has a much 
more global pattern of meridional flow with usually only a single cell
between equator and pole \citep{letal10}. Mars \citep{hetal11} and Earth
\citep{l67} have two or three cells in each hemisphere. In the Earth's
atmosphere, the so-called Hadley cell is the counterpart to the primary
meridional cell in the solar convection zone. This Hadley cell, and the
so-called Ferrell cell poleward of it, which is a countercell, show
substantial variations in amplitude, latitudinal extent, as well as
multiple equilibria in simple models \citep{hzl11,ls11,ap10,bs10,flc07}. 
The driver of these circulations is different than in stars, nevertheless 
the observed patterns of planetary and solar circulation cells compare well.

As the Sun demonstrates, well defined meridional circulation exists and
persists in the solar convection zone despite the much larger rms velocities
of the convective turbulence. This turbulence, influenced by rotation, is
generally thought to be the driver of the Sun's differential rotation, and
may also play a role in maintaining the meridional flow. A good place to
start in considering a theory for meridional circulation is \citet{r05}, who
showed that in a mean field theory of differential rotation, the buildup of
higher angular velocity in equatorial latitudes leads to an outward
radial Coriolis force which drives fluid up to the outer boundary from
below. By mass conservation this fluid must flow toward the poles and eventually
return to the bottom at higher latitudes to complete the flow circuit. What
is less clear from this reasoning is to how high a latitude  should the
circulation reach. For flux-transport dynamos, this is a crucial question,
since the period of the solar dynamo may be determined by the length of
this meridional circulation 'conveyor belt' \citep {dgdu10}.

\begin{figure}[hbt]
\epsscale{0.8}
\plotone{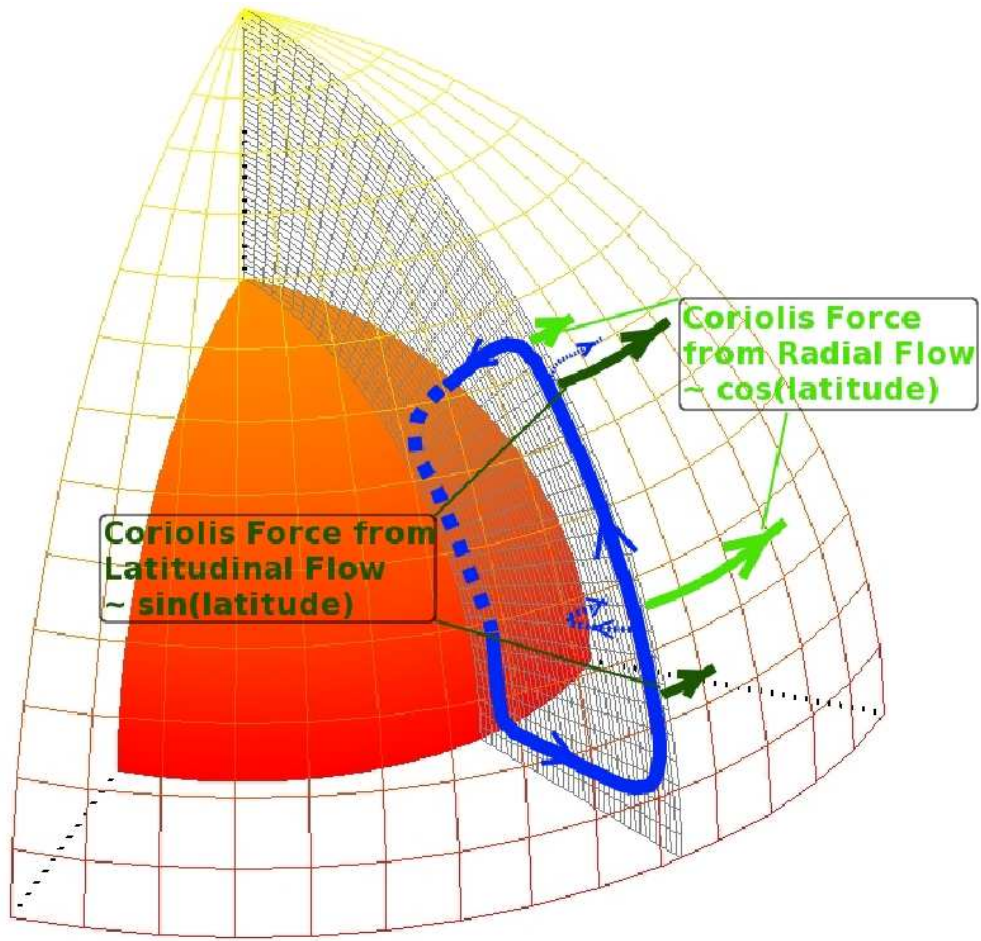}
\caption{ Schematic diagram of Coriolis forces from meridional
flow as a function of latitude, illustrating that the force from radial
flow is large in low latitudes and low at high latitudes, while the
Coriolis force from latitudinal flow is small in low latitudes and
large in high latitudes.}
\label{schematic}
\end{figure}

The radial Coriolis force may drive the meridional circulation in low
and perhaps midlatitudes, but it can not do so in high latitudes, because,
since there the rotation axis and the local vertical are nearly parallel,
the radial Coriolis force is very weak (see Figure 2). Moreover, retaining 
it in the equations of motion complicates the problem mathematically. 
In particular, retaining the radial component of Coriolis force precludes
using separation of variables to solve even the axisymmetric problem for
a spherical polar cap. Therefore we will begin from the spherical shell
equations but then approximate the spherical polar cap (see Figure 3) by 
a cylinder with the same polar axis; this eliminates the relatively mild 
spherical curvature in the cap while retaining the strong convergence to 
the pole. Then the outer boundary of the cylinder is identified with low 
latitude boundary of the spherical polar cap. 

\begin{figure}[hbt]
\epsscale{0.7}
\plotone{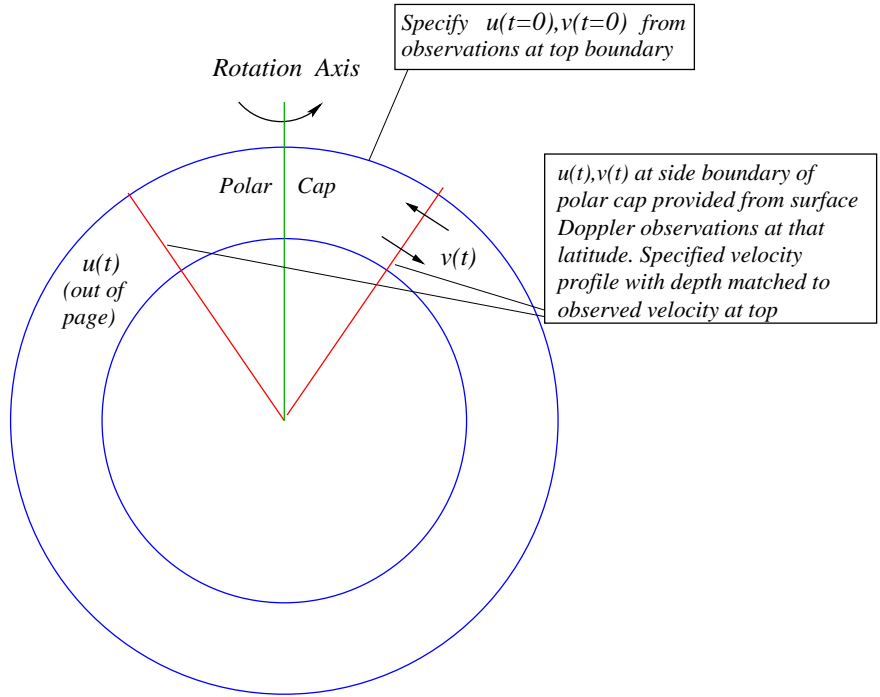}
\caption{Schematic diagram showing our 'polar cap' computation
domain for calculating meridional flow and differential rotation. 
This domain is bounded in latitude by the red radial
lines, and at the top and bottom boundaries of the convection zone,
depicted by the concentric blue circles. For simulations of time average
meridional flow and differential rotation in the polar cap, both 
velocities would be specified on the latitudinal boundary, guided by surface
observations at that latitude. These solutions are compared to the 
observed surface flows at the top of the cap. For simulations of time 
varying flow, the model would use time dependent side boundary conditions, 
again taken from observations, and would initialize the integration
with observations at the top boundary of the cap.}
\label{sphericalschem}
\end{figure}

We recognize that imposing such a boundary is somewhat artificial, since no 
such physical boundary exists in the Sun. But we can think of it as the 
location where the physics that determines the meridional flow changes from the forces that are dominant in low and mid-latitudes, to those that are most 
important at high latitudes. We do not know exactly the latitude where the 
dominant forces change, so we treat the location of the boundary as a parameter
of the problem. One guide to its placement comes from numerical simulations
of global convection in deep rotating spherical shells \citep{g79,mbdt08}
that show for a shell of the thickness of the solar convection zone as a 
fraction of the radius, the latitudinal Reynolds stress that transports
angular momentum toward the equator to maintain the differential rotation
reaches only to about $60^{\circ}$ latitude.

But even this cylindrical problem must be solved numerically; we can get a 
first idea of the nature of meridional flow in high latitudes by further 
approximating the cylinder with a cartesian analog. In this analog the 
cylinder is replaced by a channel infinite in longitude, whose left side 
boundary is identified with the axis of the cylinder, and whose right side 
boundary is identified with the outer wall of the cylinder. Thus through this 
sequence of transformations of the equations we can trace the side boundaries
of the cartesian channel back to the polar axis and equatorward boundary of 
the spherical polar cap. This cartesian geometry allows solutions in 'latitude'
in terms of simple periodic and exponential functions. We judge that such a
simplification is justified for a first study of the polarcell problem,
but should be followed by much more realistic systems, which we plan for future
papers.

To make this connection back to the spherical problem as strong as possible,
we apply the same boundary conditions in the cartesian case as we would have 
in the cylindrical and spherical cases. In words, in the spherical cap case 
these conditions are that all physical variables remain bounded at the polar 
axis and axisymmetry is maintained. These requirements imply that the 
azimuthal flow, the latitudinal flow, the latitudinal pressure gradient 
and the viscous stress all vanish on the axis. To actually solve the
cylindrical problem, we would apply the same conditions on the axis of the
cylinder, but latitude is replaced by the cylindrical radius variable
as a coordinate. In the cartesian analog, the cylindrical radius variable is
replaced by the cross-channel coordinate of the infinite channel, and the
azimuthal coordinate by the coordinate along the channel. In all three
systems, the meridional and azimuthal flows are specified on the boundary
that corresponds to the equatorward boundary of the spherical polar cap,
namely the outer boundary of the cylinder, and the right hand side boundary
of the channel. 

In the cartesian problem, the boundary conditions at the sides
introduce the possibility that momentum associated with the flow parallel
to the channel walls can enter the domain from the right side and exit from
the left, which has no counterpart in the cylindrical or spherical cases.
But we can show that this transfer has no effect on the solutions for 
meridional flow, so it is not significant dynamically.

In determining meridional circulation quantitatively, the large increase in
fluid density with depth in the convection zone has to be important, but it
also adds considerable mathematical complexity. It is possible to write
the equations of motion in terms of mass flux rather than velocities, in
which case the effect of the density increase with depth is seen explicitly 
only in lower order terms in the turbulent diffusion expressions. We will
keep these terms in the initial equations shown below, but set aside
this variation in looking for the first simplest solutions.  

Other effects of possible importance in determining meridional circulation
in the solar convection zone include departures from the adiabatic gradient,
so that buoyancy forces come into the problem, and jxB forces from magnetic
fields. Departures from the adiabatic gradient, particularly their 
variations in latitude, could be important for determining the profile of 
differential rotation in the bulk of the convection zone and at its base, but 
the meridional flow is not very different from the case with adiabatic 
stratification. Therefore we will initially leave out such departures
from the adiabatic. This means that when the hydrostatic pressure of the 
reference state is subtracted from the equations, gravity (which includes the
centrifugal force of the rotating coordinate system) drops out of the 
problem. The effects of magnetic stresses are also deferred to a later paper.
Because meridional flow and differential rotation are very subsonic flows,
we do not expect acoustic effects to be important, so we start from equations
in which acoustic modes have been filtered out. In that case the 
mass-continuity equation contains no time derivative. We are in effect using 
the so-called 'anelastic' equations.

\section{MATHEMATICAL FORMULATION}

In this section we develop the  hierarchy of equations for meridional
circulation and differential rotation in the polar cap, starting from 
vector-invariant forms of the equations of motion and mass conservation.  

\subsection{Vector invariant and spherical component equations}

In vector form, the equations of motion are given by

$${\partial {\v} \over \partial t} = -{1 \over \rho}\grad p - \grad
\left(\v\cdot \v/2 \right) + \v \vcross\left(\grad \vcross \v \right)
- 2 {\Omega} \vcross \v-{1 \over \rho}\grad \vcross \rho \nu
\left(\grad \vcross\v \right)- g \hat{z} \quad\eqno(1)$$
and
$$\Div\left(\rho \v \right)=0.\quad\eqno(2)$$

If we subtract out a reference state hydrostatic balance, then gravity is
removed from the problem and the pressure variable $p$ contains the departure
of pressure from this hydrostatic balance. In this anelastic system, the
density $\rho$ is the reference state density. Then in spherical coordinates
($r,\theta,\phi$) the component equations for the flow are

$${\partial w \over \partial t}=-{1 \over \rho}{\partial p \over
\partial r}+2\Omega\sin\theta u-{1 \over {\rho r^2\sin\theta}}{\partial
\over \partial \theta}\left(\rho \nu \sin \theta\left({\partial \over \partial
r}(rv)-{\partial w \over \partial \theta}\right)\right)+F_r ; \quad\eqno(3)$$
$${\partial v \over \partial t}=-{1 \over \rho r}{\partial p \over
\partial \theta}+2\Omega\cos\theta u +{1 \over \rho r}{\partial \over \partial
r}\left(\rho \nu \left({\partial \over \partial r}(rv)-{\partial w \over 
\partial \theta}\right)\right)+F_{\theta};\quad\eqno(4)$$
$${\partial u \over \partial t}=-2\Omega\cos\theta v-2\Omega\sin\theta w
+{1 \over \rho r}\left({\partial \over \partial r}\left(\rho \nu 
{\partial \over \partial r}(ru)\right)+{1 \over r}{\partial \over
\partial\theta}\left({\rho \nu \over \sin\theta}{\partial \over \partial 
\theta} \left(\sin\theta u\right)\right)\right)+F_{\phi}.\quad\eqno(5)$$

These equations are subject to the constraint of mass-conservation,
given by
$${1 \over r}{\partial \over \partial r}(\rho r^2 w)+{1 \over \sin\theta}
{\partial \over \partial \theta}(\rho \sin \theta v)=0.\quad\eqno(6)$$

In these equations, $u$ is the (linear) rotational velocity, $v$ the
velocity in colatitude $\theta$ and $w$ the radial velocity. $\Omega$ is
the rotation rate of the coordinate system, $\rho$ the reference state 
density, which varies only in radius, $p$ the pressure relative to a spherically
symmetric hydrostatic pressure, $\nu$ the turbulent viscosity.
$F_r, F_{\theta}$ and $F_{\phi}$ are force components that would arise
from turbulent Reynolds stresses. Nonlinear inertial terms involving
products of axisymmetric meridional and rotational flow have been
excluded from equations for simplicity and because the turbulent
Reynolds stresses should be larger. A nonzero $F_{\phi}$ in the polar
cap leads to `gyroscopic pumping' as described in \citet{m07}. In particular,
if this process is at work, in order to get poleward flow the Reynolds stress
must be transporting angular momentum toward the equator there, by an amount
that increases with colatitude. 

\subsection{Spherical equations for streamfunction and rotational velocity}

Equations (3)-(6) above can be solved as is, or the perturbation pressure
can be eliminated by cross-differentiation of equations (3) and (4), together
with defining a streamfunction $\chi$ for the meridional flow $v,w$ such that
the mass continuity equation (6) is satisfied automatically. By inspection,
this streamfunction is defined so that

$$ \rho w={1 \over r \sin \theta}{\partial \over \partial \theta}
( \sin \theta \chi); 
\rho v=-{1 \over r}{\partial \over \partial r}
(r \chi).\quad\eqno(7)$$

The prediction equation for $\chi$ is given by

$${\partial \over \partial t}L \chi=-{2\Omega \cos \theta \over r}{\partial 
\over \partial r}(r \rho u)+{2\Omega \over r} {\partial \over \partial \theta}
(\sin \theta \rho u)$$
$$+ L(\nu L \chi) - L\left({\nu \over r}{\partial (ln\rho) \over \partial r}
{\partial \over \partial r}(r \chi)\right)$$ 
$$-{1 \over r} {\partial \over \partial r} (\rho rF_{\theta})+{\rho \over r} 
{\partial \over \partial \theta}F_r,\quad\eqno(8)$$

in which the $L\chi$ is defined by

$$L\chi=\left(\nabla^2- {1 \over r^2 \sin^2 \theta}\right)\chi.\quad\eqno(9)$$

Here $\nabla^2$ is the standard axisymmetric Laplacian operator in spherical 
polar coordinates.

Finally the prediction equation (5) for the rotational flow $u$ becomes

$${\partial \over \partial t}u={2\Omega \cos \theta \over \rho r}{\partial
 \over \partial r}(r \chi)-{2 \Omega \over \rho r}{\partial \over \partial
\theta}(\sin \theta \chi)+{1 \over \rho r}{\partial \over \partial r}
(\rho \nu {\partial \over \partial r}(ru))+{1 \over \rho r^2}{\partial
\over \partial \theta}({\rho \nu \over \sin \theta} {\partial \over \partial 
\theta} (\sin \theta u))+ F_{\phi}. \quad\eqno(10)$$

Equations (8)-(10) can be solved numerically. However, this is the first
attempt to understand the flow behavior in the Sun's polar latitudes. So, in 
order to understand how this model works, we seek further approximations to a 
cylindrical system as an intermediate step before going to a rectangular system. 

\subsection{Cylindrical component equations}

In this subsection, we define a cylindrical coordinate system whose axis 
coincides with the rotation axis of the Sun. The independent variables in 
this system are $s,\phi,z$; $s$ is the outward radial coordinate (colatitude 
at the pole), $\phi$ the azimuthal coordinate (longitude on the Sun), and $z$ 
the axial coordinate (corresponding to the outward radial coordinate at the 
pole). The corresponding velocities for these coordinates are defined as 
$v,u,w$. If we restrict ourselves to axisymmetric variables, then the azimuthal
flow $u$ is identified with the solar differential rotation linear velocity
in high latitudes, and $v,w$ with the solar meridional circulation there. 
In the axisymmetric case, the three component equations of fluid motion are 
given by,

$$ {\partial v \over \partial t}= -{1 \over \rho}{\partial p \over \partial s}
-{\partial \over \partial s}\left({v^2+u^2+w^2 \over 2}\right)+{u \over
s}{\partial \over \partial s}\left(su\right)+w\left({\partial w \over \partial
s}-{\partial v \over \partial z}\right)$$
$$ +2\Omega u-{1 \over {{\rho}s}}{\partial \over \partial z}
\left(s\rho\nu\left({\partial w \over \partial s}-{\partial v \over 
\partial z}\right)\right);\quad\eqno(11)$$

$${\partial u \over \partial t}=-{v \over s}{\partial \over \partial s}(su)
-w{\partial \over \partial z}(su)-2\Omega v+{1 \over \rho}{\partial
\over \partial s}\left({\rho\nu \over s}{\partial \over \partial
s}(su)\right) + {1 \over \rho}{\partial \over \partial z}\left({\rho\nu \over
s}{\partial \over \partial z}(su)\right);\quad\eqno(12)$$

$${\partial w \over \partial t}=-{1 \over \rho}{\partial p \over
\partial z}-{\partial \over \partial z}\left({v^2+u^2+w^2 \over
2}\right)+u{\partial \over \partial z}(su)-v\left({\partial w \over 
\partial s}-{\partial v \over \partial z}\right)$$
$$+{1 \over \rho s}{\partial \over \partial s}\left(s\rho\nu\left({\partial w
\over \partial s}-{\partial v \over \partial z}\right)\right).\quad\eqno(13)$$

In cylindrical coordinates the axially symmetric mass continuity equation
is given by,

$${1 \over s}{\partial \over \partial s}(\rho sv)+{\partial \over \partial z}
(\rho w)=0.\quad\eqno(14)$$

\subsection{Steady state linear equations}

The simplest meaningful problem we can solve is one in which the flow is
taken to be steady and the velocities relative to the rotating reference
frame are small enough that the nonlinear inertial terms can be dropped.
We also take $\rho$ and $\nu$ to be independent of $s$ and functions only
of $z$. The fluid flow inside the cylinder is forced by fluid entering and 
leaving from the outer radial boundary (the equivalent of a latitude circle 
in the spherical case).

With the above conditions, the mass continuity equation (14) is unchanged,
while the equations of motion (11)-(13) reduce to

$${1 \over \rho}{\partial \over \partial z}\left(\rho\nu\left({\partial
w \over \partial s}-{\partial v \over \partial z}\right)\right)+{1\over
\rho}{\partial p \over \partial s}-2\Omega u=0;\quad\eqno(15)$$

$$\nu{\partial \over \partial s}\left({1 \over s}{\partial \over
\partial s}(su)\right)+{1 \over \rho}{\partial \over \partial
z}\left(\rho\nu{\partial u \over \partial z}\right)-2\Omega v=0;\quad\eqno(16)$$

$${\nu \over s}{\partial \over \partial s}\left(s\left({\partial w \over
\partial s}-{\partial v \over \partial z}\right)\right)-{1 \over
\rho}{\partial p \over \partial z}=0.\quad\eqno(17)$$

The physical boundary conditions we take for the cylinder are to allow
no flow through, and no viscous stress on, the top and bottom; we impose
meridional flow on the outer radial boundary, representing the primary 
poleward cell in the Sun. To obtain steady solutions, there must also be
no net torque applied to the outer boundary. As stated above, certain 
conditions must also be imposed on the axis, to ensure all physical variables 
remain bounded and single-valued there and axisymmetry is maintained. These 
requirements imply that azimuthal flow $u$, radial flow $v$, radial pressure 
gradient and viscous stress all vanish there.

\subsection{Separation of variables: incompressibile case}

\subsubsection{Separated equations}

If we discard density variations in the $z$ direction, so density is
constant, we can absorb the density into the pressure, defining a new variable
$\Pi$=$p/\rho$. Then we define a stream function $\psi$ that satisfies
the mass continuity equation through the relations

$$v=-{\partial \psi \over \partial z};w={1 \over s}{\partial \over
\partial s} (s\psi).\quad\eqno(18)$$

With these variable changes, equations (15)-(17) become

$$\nu{\partial \over \partial z}\left(\nabla^2
\psi-\psi/s^2\right)-2\Omega u+{\partial \Pi \over \partial
s}=0;\quad\eqno(19)$$

$$\nu\nabla^2 u+2\Omega {\partial \psi \over \partial z}=0;\quad\eqno(20)$$

$${\nu \over s}{\partial \over \partial s}\left(s(\nabla^2
\psi-\psi/s^2)\right) -{\partial \Pi \over \partial z}=0.\quad\eqno(21)$$

In equations (19)-(21), $\nabla^2$ is the standard axisymmetric cylindrical
Laplacian operator.

We can eliminate the pressure variable $\Pi$ from equations (19) and (21)
by cross differentiation, reducing the system to one second order equation,
equation (20), and one fourth order equation, given by

$$\left(\left(\nabla^2-1/s^2\right)\left(\nabla^2 \psi-\psi/s^2\right)\right)
-2\Omega {\partial u \over \partial z}=0.\quad\eqno(22)$$

There are two distinct classes of solutions to the system (20),(22). Since
these equations are linear, these solutions can be superimposed with
relative amplitudes determined by the boundary conditions for each. One class
is found by setting the meridional flow streamfunction $\psi=0$ everywhere. It
yields pure differential rotation independent of $z$ forced at the outer 
boundary of the cylinder. To satisfy the condition that there be no net
torque at the outer boundary, to allow steady solutions in the interior of
the cyclinder, there must be no viscous stress at this boundary. This 
corresponds to solutions with constant angular velocity, or linear
rotational velocity $u$ that decreases linearly with $s$ toward the axis of
the cylinder. The other class of solutions can be found by separation of 
variables, since coefficients in this system are functions of $s$ only. We 
place the lower and upper boundaries of our cylinder at $z=0,H$ respectively. 
Then if we allow no flow through the lower or upper boundary, these boundaries 
must coincide with a streamline, and there should be no viscous stress there 
either. These conditions are satisfied if we take

$$\psi=\sum_{n=1}^{\infty}\Psi_n \sin(n\pi z/H); u=\sum_{n=1}^{\infty}
U_n \cos(n\pi z/H).\quad\eqno(23)$$

With this choice of representation of the solutions, we can find separate
solutions for each $n$ for $\psi$ and $u$. We can then represent the forcing
at the boundary in the same way, so the amplitude of the solutions for each
$n$ is determined separately by the amplitude of the forcing for the same $n$.
There could be other solutions that use other representations, but we have
not looked for them.

Then if we substitute expressions (23) into equations (20),(22), and define
$\sigma_n=n\pi/H$, then equations (20) and (22) become, for each $n$,
$$\nu {{\nabla}^2}_n U_n +2\Omega \sigma_n \Psi_n=0,\quad\eqno(24)$$
and
$$\nu(\left(\nabla_n^2-1/s^2\right)\left(\nabla_n^2
\Psi_n-\Psi_n/s^2\right)-2\Omega \sigma_n U_n=0,\quad\eqno(25)$$
in which
$$\nabla_n^2={1 \over s}{d \over ds}s{d \over ds}-\sigma_n^2.\quad\eqno(26)$$

The set of solutions to equations for which there is no meridional flow
are for $u$ only, which from equation (22) must satisfy the relation
${\partial u \over \partial z}=0$ everywhere, so $u$ is a function of $s$
only; $u$ is independent of $z$. For $\psi=0$ everywhere, equation (20) 
becomes ${\partial \over \partial s}(s{\partial u \over \partial s})=0$. 
The solutions to this equation all are for constant angular velocity, or
linear rotational velocity that is a linear function of $s$. Since this 
linear velocity is independent of $z$ it will also satisfy stress free boundary
conditions at top and bottom. But since $\psi=0$ in these solutions, they do 
not contribute to determining the amplitude of meridional flow or where the 
nodes in this flow occur. These properties are determined entirely from 
solutions to equations (24)-(26) for $\Psi_n$ and $U_n$. The solutions for 
$u$ with $\psi=0$ everywhere can be used to match the total solution for 
differential rotation linear velocity with observational estimates. These total
solutions will therefore have a differential rotation that is in part 
independent of $z$ and in part a cosine function of $z$.

\subsubsection{Nodimensionalization and setting of parameters}

If we scale all lengths in the problem by $H$, the depth of the
cylinder, and recognize that $U_n$ scales differently from that of 
$\Psi_n$ by $H$, then the dimensional system (24)-(25) can be written 
in dimensionless form as

$$\nabla_n^2U_n+\epsilon_n \Psi_n=0, \quad\eqno(27)$$
and
$$\left(\nabla_n^2-1/s^2 \right)^2\Psi_n-\epsilon_n U_n=0, \quad\eqno(28)$$
in which $\epsilon_n$ is a dimensionless parameter, defined as
$\epsilon_n =2\Omega\sigma_n H^3/\nu$. In this dimensionless system, velocities
are scaled by $\nu/H$ and therefore the streamfunction by $\nu$.

For this system the boundary conditions, defined physically in section 3.3,
that should be applied are that

$$\Psi_n=0; \,\,{d^2 \Psi_n\over ds^2}=0; \,\, U_n=0; \,\, {d U_n\over ds}=0 
\,\,\, {\rm at} \,\,\, s=0; \quad\eqno(29a)$$
and
$$ \Psi_n \,\, {\rm and} \,\, U_n \,\,\, {\rm specified \,\,\, at} \,\,\,
s=R .\quad\eqno(29b)$$

These boundary conditions completely specify the system for solutions
containing both meridional flow and differential rotation for which $n\geq 1$.
For forcing with any $n$ imposed on the boundary, there will be a response
in meridional circulation and differential rotation in the interior that 
contains the same $n$. Since the problem is linear, solutions with multiple
$n$ values can be constructed simply by adding together solutions for each
individual $n$. The relative amplitudes of each of these solutions would be
determined by the relative amplitudes of the forcing for different $n$.

If equations (27) and (28) were combined into a single equation for $\Psi_n$,
it would be sixth order, yielding a total of six independent solutions. We 
would need to apply a total of six boundary conditions; these include four on 
the $z$ axis, and two at the outer radial boundary. For any added differential
rotation component $U_0$ that satisfies the equation $d(s(d U_0/ds))/ds=0$
the boundary conditions are $U_0=0$ and $dU_0/ds=0$ at $s=0$, and $U_0$
specified at $s=R$.

$\epsilon_n$ is the only dimensionless parameter that is explicit in the 
equation system (27)-(29), but in fact there are others, implied
or implicit in this system, which may be useful in evaluating the results.
We define them here, and discuss their possible significance. 

A traditional dimensionless number used to characterize the relative influence
of Coriolis and viscous forces in a fluid dynamical problem is the so-called
Taylor number $Ta=4\Omega^2 H^4/\nu^2$. By inspection, it is related to
$\epsilon_n$ by the relationship $\epsilon_n=n\pi {Ta}^{1/2}$. In a typical 
problem involving both rotation and viscosity, $Ta$ needs to be $10^3$ or
higher to show substantial influence of rotation. If we take $\nu\sim 10^{12}
{\rm cm}^2\,{\rm s}^{-1}$, $Ta=4 \times 10^6$ for the whole depth of the solar 
convection zone, so we should expect a substantial influence of rotation on 
the solutions of most interest for the Sun. Another number often used in such 
problems is the Ekman number $E=1/2Ta^{1/2}$. This number would be small for
the solar convection zone; it measures the thickness of the boundary layers
that would form if the top or bottom of the convection zone were considered
to be 'nonslip'. But, more realisticly, we take these boundaries to be stress
free, so no Ekman layers are allowed to form.

Because the rotational velocity $U$ is specified on the outer boundary of
the cylinder, we can also define a Rossby number $Ro=U/2 \Omega H$ and a 
Reynolds number $Re=UH/\nu$. A plausible linear rotational velocity
at the outer boundary would be $60 {\rm m}\,{\rm s}^{-1}$, for which 
$Ro=0.115$. When $Ro<<1$ the flow tends to be geostrophic; that is, there 
is a near balance in latitude (radius in the cylinder) between Coriolis 
and pressure gradient forces. This balance should therefore be checked in 
the solutions we find. With the scaling we have used, the Reynolds number 
is the velocity itself. We should expect different behavior of the solutions 
depending on whether $Re>>1$ or $Re<<1$. We will see that we can get $Re>>1$ 
in our model only if we assume quite low turbulent viscosity $\nu \sim 10^{10}
{\rm cm}^2\,{\rm s}^{-1}$. For most of the parameter space of interest $Re<1$, 
in many cases much less. In general, the implication of small $Re$ is that 
the flows are likely to be smooth and laminar, i.e., not turbulent. If the 
flow imposed at the low latitude boundary of the polar cap (outer boundary 
of the cylinder) is time dependent, the interior flows induced will also be, 
but will generally have similar time dependence to that seen at the boundary.

\subsection{Formulation in cartesian geometry}

\subsubsection{Equations}

We can get a preliminary idea of the nature of the solutions for meridional
circulation and differential rotation in the cylindrical problem if we
look at it in a 'cartesian limit', found by conceptually replacing the full 
cylinder by a cylindrical annulus and taking both the inner and outer 
boundaries of the annulus to a large distance from the axis. 
Then all curvature effects can be ignored; a disadvantage is that we lose
the convergence of the meridional flow to the polar axis. The effect of this
approximation will need to be assessed in a later study. In this cartesian
limit, equations (27) and (28) reduce to

$$\left({d^2 \over dx^2}-\sigma_n^2\right)U_n+\epsilon_n
\Psi_n=0\quad\eqno(30)$$
and
$$\left({d^2\over dx^2}-\sigma_n^2\right)^2\Psi_n-\epsilon_nU_n=0,\quad\eqno(31)
$$
in which we have replaced $s$ with $x$ as the independent variable to avoid
confusion between cylindrical and cartesian coordinates. Here the quantity 
$\sigma_n=n \pi$ is the dimensionless form.

Equations (30) and (31) admit of solutions of the form

$$\left(\Psi_n,U_n\right) \sim e^{k_nx},\quad\eqno(32)$$
in which $k_n$ is in general complex. Substitution of the forms (32) into
equations (30) and (31) yields a $3\times 3$ determinant of the coefficients, 
which must vanish for there to be a nontrivial solution to the equations. 
This yields the relationship

$$\left(k_n^2-n^2\pi^2\right)^3+\epsilon_n^2=0.\quad\eqno(33)$$

This equation has six distinct roots in $k_n$ for each $n$; they come in two
sets of three, from

$$k_n=\pm \left(n^2
\pi^2+(-1)^{1/3}\epsilon_n^{2/3}\right)^{1/2}.\quad\eqno(34)$$

The argument of the square root has three distinct values, because
$-1 =e^{\pm i\pi},e^{\pm 3i\pi}$ and therefore $(-1)^{1/3}=e^{\pm
i\pi/3},-1$. The first two choices yield complex $k_n$, meaning that 
these solutions are both exponential and sinusoidal with respect to $x$, 
while the third choice leads to $k_n$ either real or imaginary, depending 
on the values of $n$ and $\epsilon_n$. These solutions are therefore 
either purely exponential or purely oscillatory with respect to $x$. 
In general, we expect to need all six solutions for $k_{np}, p=1,6$ to 
satisfy the boundary conditions at the sides of the Cartesian channel. 

As stated above, the left hand edge of the cartesian channel corresponds 
to the axis in cylindrical geometry. In the cylindrical case we required 
$U_n=0$ on the axis, to keep the angular velocity finite and avoid 
multiple-valued functions, but, strictly speaking, these conditions do not 
apply in the cartesian case. However, for comparison with the cylindrical and
spherical systems, we retain the same boundary conditions. This has the 
artificial effect of allowing momentum to cross this boundary, which can not
happen in the cylindrical or spherical cases. But this property is
inconsequential in that it has no effect on the solutions for meridional
circulation.

In the cartesian limit we have distorted only the geometry in order to find 
analytical solutions that should tell us something about how extended the 
primary meridional flow cell is as a function of the parameters of the problem,
particularly $\epsilon_n$, which is the same in cartesian and cylindrical cases.
Therefore in cartesian coordinates the boundary conditions are

$$\Psi_n=0,{d^2\Psi_n \over d^2x}=0,U_n=0,{dU_n \over dx}=0, \,\,\,
{\rm at} \,\,\, x=0;\quad\Psi_n,U_n \,\,\, 
{\rm specified \,\, at} \,\,\, x=R. \quad\eqno(35)$$

Next we generalize the equation forms (32) to account for the six solutions for
each variable by representing each as

$$ \Psi_n = \sum_{p=1}^6 \Psi_{np}e^{k_{np}x},
\quad U_n=\sum_{p=1}^6 U_{np}e^{k_{np}x}. \quad\eqno(36)$$

Then satisfaction of the boundary conditions at $x=0$ requires that

$$ \sum_{p=1}^6 \Psi_{np} = 0, \quad \sum_{p=1}^6 k_{np}^2\Psi_{np} = 0, 
\quad \sum_{p=1}^6 U_{np}= 0, \quad \sum_{p=1}^6k_{np} U_{np}=0. 
\quad\eqno(37)$$

Satisfaction of the inhomogeneous boundary conditions at the right hand
boundary of the channel (corresponding to the outer radial boundary of
the cylinder) requires

$$\sum_{p=1}^6 \Psi_{np}e^{k_{np}R} = \Psi_n(R), \quad \sum_{p=1}^6
U_{np}e^{k_{np}R}=U_n(R), \quad\eqno(38)$$
in which $\Psi_n(R)$ and $U_n(R)$ are the values imposed on the system
by the boundary conditions at the right hand boundary $x=R$.

Then equations (37) and (38) are a system of six linear equations, four
homogeneous, two inhomogeneous, for 12 independent variables -- seemingly
a very underdetermined system. But we must recognize that equations (30)-(31)
impose relationships among $\Psi_{np},U_{np}$, since they all must be
satisfied separately for each $n,p$. These relationships are

$$(k_{np}^2-\sigma_n^2)U_{np}+\epsilon_n\Psi_{np}=0;\quad\quad p=1,6
\quad\eqno(39)$$
and
$$(k_{np}^2-\sigma_n^2)^2\Psi_{np}-\epsilon_nU_{np}=0;\quad\quad p=1,6
.\quad\eqno(40)$$

If we invoke all these relationships, then it would seem that the system is
overdetermined, but since the $k_{np}$ are found from equation (33) also by 
substitution of form (32) into (30)-(31) only one of equations (39)-(40) are in
fact independent. Therefore (39)-(40) provide twelve independent relationships
among $\Psi_{np}$ and $U_{np}$, exactly what we need to constrain
and solve the system (37)-(38) for these variables. In practice, we
eliminate $U_{np}$ from (37) and (38) by substitution from (39). These 
equations are:

$$ \sum_{p=1}^6 \Psi_{np}=0;\quad\eqno(41)$$

$$\sum_{p=1}^6 {k_{np}^2}\Psi_{np}=0;\quad\eqno(42)$$

$$\sum_{p=1}^6{1\over(k_{np}^2-\sigma_n^2)}\Psi_{np}=0;\quad\eqno(43)$$

$$\sum_{p=1}^6{k_{np}\over (k_{np}^2-\sigma_n^2)}\Psi_{np}=0;\quad\eqno(44)$$

$$\sum_{p=1}^6 e^{k_{np}R}\Psi_{np}= \Psi_n(R); \quad\eqno(45)$$

and

$$\sum_{p=1}^6 {\epsilon_n e^{k_{np}R}\over
(k_{np}^2-\sigma_n^2)}\Psi_{np}=-U_n(R).\quad\eqno(46)$$

These equations can be solved using Cramer's rule, provided the $6\times 6$
determinant of the coefficients of these equations does not vanish. This
system has either one or two inhomogeneous equations, and respectively
either five or four homogeneous equations.

By separating the real and imaginary parts of these six complex
equations, we numerically solve the system of twelve linear equations 
by factoring into lower and upper triangular form followed by back 
substitution. The magnitudes of solution vectors varies widely in the
range of parameters of our interest. However, we have found that it is 
always the case that the highest magnitude of $L^2 = || {\bf A}\cdot{\bf x} 
- {\bf b} ||$ norm error of the solved system is at least twelve orders 
of magnitude lower than the magnitudes of solution vectors.

If we wish to add to these a solution for pure differential rotation
(no meridional flow) driven by a linear rotation velocity independent of
$z$ on the right hand channel boundary, it must satisfy the relationship
$d^2U_0/d^2x=0$ with $U_0=0$ at $x=0$ and $U_0$ specified at the right
hand boundary, so this flow simply decreases linearly to zero from right
to left, corresponding to constant angular velocity in the cyclindrical
and spherical cases. Not surprisingly, in the cartesian case, it is not possible
to satisfy the boundary condition $dU_0/dx=0$ at $x=0$, because momentum
that flows in from the right side must flow out on the left side to
maintain a steady state. As stated above for the cylindrical case, these
linear solutions have no effect on the amplitudes or the positions of the nodes
in $x$ for meridional flow or differential rotation that are functions of $z$. 

\subsubsection{Parameter ranges of interest}

There are four independent parameters to be varied in our dimensionless 
cartesian system. These include $\epsilon_n$, which measures the relative
influence of Coriolis and viscous forces; $R$, which measures the aspect
ratio of the channel and therefore the colatitude of the equatorial boundary
of the polar cap; $U_n(R)$, the linear rotational velocity imposed at this
boundary; and $\Psi_n(R)$, representing the meridional flow imposed there.
To estimate the range of $\epsilon_n$ of interest, we use for $\Omega$ the
core rotation rate of the Sun, $2.6\times 10^{-6} {\rm s}^{-1}$ and for $H$ 
the depth of the convection zone, $2 \times 10^{10} {\rm cm}$. 
Then the range of $\epsilon_n$ is given by $6.53 \times 10^{15}n/\nu$. 
Therefore for the range $10^{11}<\nu<10^{16} {\rm cm}^2\,{\rm s}^{-1}$ for 
$n=1$, the lowest vertical mode, we get $6.53 \times 10^4>\epsilon_1> 0.653$. 
The most plausible values for $\nu$ in the convection zone of the Sun are in 
the range $10^{12}-10^{14} {\rm cm}^2\,{\rm s}^{-1}$.

A typical dimensional meridional flow speed observed in midlatitudes near the 
boundary of our polar cap falls in the range -5 to -20 ${\rm m}\,{\rm s}^{-1}$ 
(the negative sign is for flow toward the left hand edge of the channel,
corresponding to poleward flow in the polar cap). Relative to the core rotation 
rate, a surface linear rotational flow speed at similar latitudes would be in 
the range -20 to -90 ${\rm m}\,{\rm s}^{-1}$. What dimensionless
values these correspond to vary according to what value of $\nu$ we assume.
For $5\,{\rm m}\,{\rm s}^{-1}$ and $\nu$ of $10^{11}\, {\rm cm}^2\,{\rm s}
^{-1}$ we get a dimensionless velocity of one unit and a streamfunction of 
$1/\pi$ units. For a dimensional speed of $90\, {\rm m}\,{\rm s}^{-1}$ and 
$\nu$ of $10^{15} {\rm cm}^2\,{\rm s}^{-1}$, the dimensionless velocity is 
$1.8 \times 10^{-4}$ units. The streamfunction associated with a peak 
meridional flow of $20\,{\rm m}{\rm s}^{-1}$ and $\nu=10^{15}\, {\rm cm}^2\,
{\rm s}^{-1}$ is $4 \times 10^{-4}/\pi$. The results that follow will be for 
parameter values within these ranges and for even higher viscosity, in order 
to show the full range of behavior of the solution with respect to 
node number and location.
 
In the case of the differential rotation linear velocity imposed at the side
boundary, we must decide the relative amplitudes of the $z$ independent part
$U_0$ and the $z$-dependent part $U_n$. This choice will be guided by
observations. Whatever the choice, only the $z$ dependent part of $U$ has
dynamical consequences for the meridional flow and the position of its nodes.

\section{RESULTS}

Our results are of two types: displays of the latitude positions of all
the nodes in the streamfunction as functions of the parameters of the problem,
and contour plots of that streamfunction. We discuss node position first.
We focus on solutions for which the boundary forcing is chosen with $n=1$,
corresponding to a primary meridional circulation cell that has poleward 
flow in the upper half of the polar cap, and equatorward flow in the lower
half. To produce a meridional flow in the polar cap that contains $n>1$
requires forcing at the boundary that includes components with $n>1$, which,
if the amplitude of the $n>1$ components were large enough, would imply the 
existence of more than one meridional flow cell with depth at all latitudes,
as well as latitudinal differential rotation changing sign with depth. We 
are not aware of observational evidence for such flows, so we have not
considered that case. We could alter the simple $\sin\pi z$ dependence with 
depth without introducing a second cell, if we added a $\sin 2 \pi z$ term 
to the forcing with a sufficiently small amplitude factor. We have not 
explored that possibility in this paper.

\subsection{Latitude locations of nodes}

Figure 4 displays the latitude positions of streamfunction nodes for a solar 
type turbulent viscosity of $\nu=10^{13} {\rm cm}^2\,{\rm s}^{-1}$ for a 
range of plausible meridional flows and four choices of differential rotation 
imposed at the boundary, which for this case is at $60^{\circ}$ latitude. 
Figure 5 displays node position as a function of boundary meridional flow 
speed for boundaries set at $60^{\circ}$, $65^{\circ}$ and $70^{\circ}$ 
(frames a,b,c respectively), for a wide range of turbulent viscosities and 
for a plausible rotation of the solar core. We recognize that the range of 
viscosities is much wider than plausible for the sun, but in this first
study we want to illustrate the full range of behavior of our model. 
Figure 6 illustrates detailed behavior of the nodes for the boundary at 
$65^{\circ}$ in a part of the parameter space in which a change in the 
meridional flow at the boundary leads to a merging of nodes, which implies that
the number of nodes drops from two to zero. Figure 7 displays the latitude 
of the first and second nodes as functions of the turbulent viscosity for the 
different boundary placements and a typical flow speed, namely $10\, 
{\rm m}\,{\rm s}^{-1}$.

Figure 4 shows that for the parameters chosen there are two nodes within the
channel, the first located between $60^{\circ}$ and $65^{\circ}$ (solid color 
curves), the second between $76^{\circ}$ and $81^{\circ}$ (dashed curves), 
depending on the meridional flow amplitude at the boundary. The red curves are
for a vertical and latitudinal differential rotation linear velocity 
that approximates the solar profile at $60^{\circ}$; hence the label 'Normal'.
The other curves are for specified fractions of that value. We see from 
Figure 4 that the latitude of nodes is only weakly a function of differential 
rotation at the boundary, so in subsequent results presented we focus on a 
solar differential rotation. We do this also because we know from 
helioseismic measurements that the differential rotation profile changes very
little with time (torsional oscillations are only a few \% of the time average
differential rotation). In addition, the latitude spacing between the
first and second nodes, about $16^{\circ})$, is virtually independent of 
choice of differential rotation and meridional flow speed.

\clearpage
\begin{figure}[hbt]
\epsscale{1.0}
\plotone{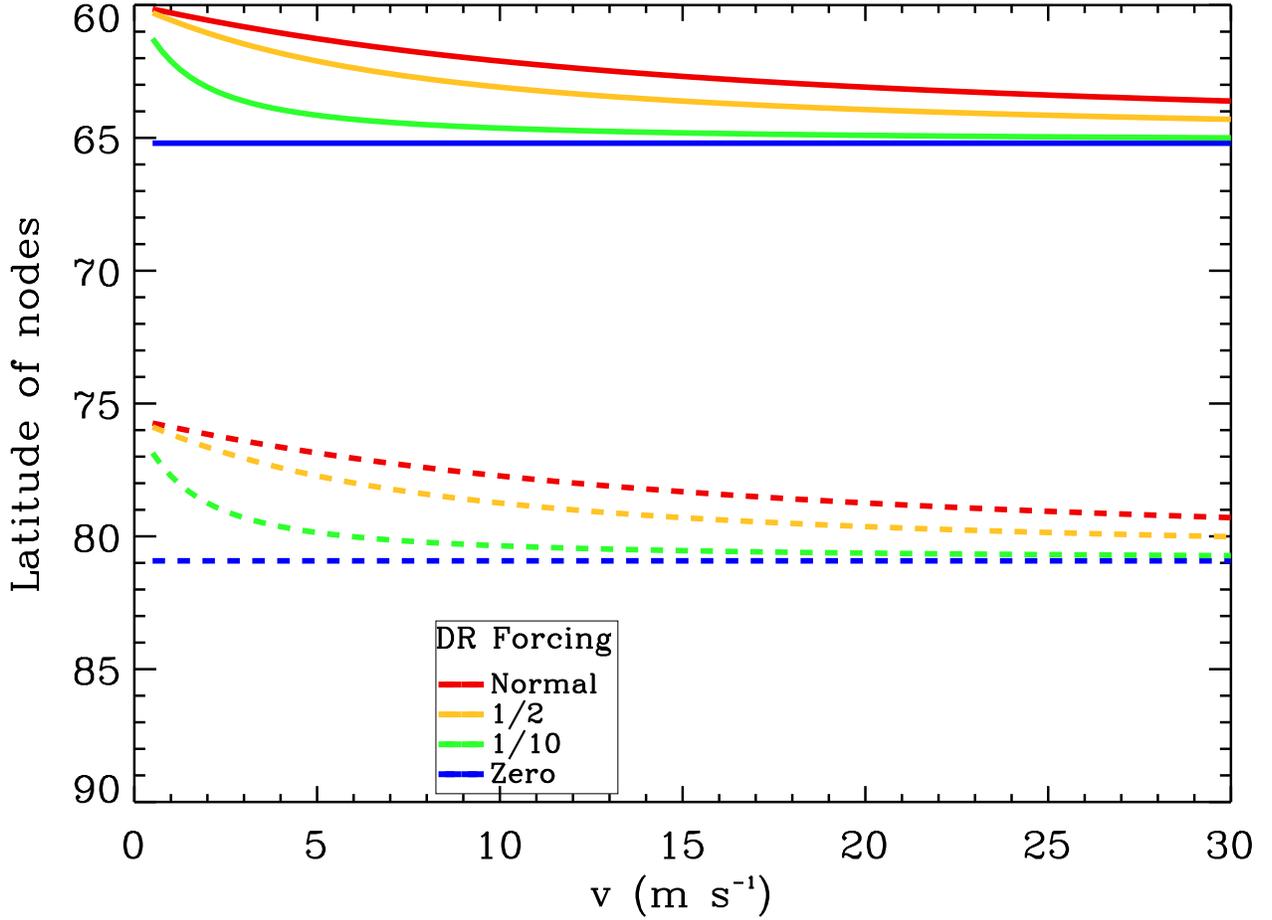}
\caption{Latitudinal positions of streamfunction nodes as a function of 
meridional flow speed at the boundary, for selected differential rotation
amplitudes at the same boundary (color key is shown in the figure). }
\label{nodeposDR}
\end{figure}

How sensitive are the node positions to the turbulent viscosity?
Figure 5 gives the answer. For all three boundary latitudes, for 
$\nu=10^{12} {\rm cm}^2\,{\rm s}^{-1}$, there are three nodes. For 
$10^{13} {\rm cm}^2\,{\rm s}^{-1}$ there are two. For $\nu=10^{15}$ 
and above there is at most one node; as the turbulent viscosity is raised, 
the range of meridional flow speeds for which there is even one node shrinks 
toward zero. For meridional flow speeds at the boundary of, say, $10\, 
{\rm m}\,{\rm s}^{-1}$, the last node does not disappear until the turbulent 
viscosity is as high as $10^{16}-10^{17} {\rm cm}^2\,{\rm s}^{-1}$, 
values two to three orders of magnitude larger than plausible for the Sun, 
even for the solar surface where supergranules are active. 

\clearpage
\begin{figure}[hbt]
\epsscale{0.47}
\plotone{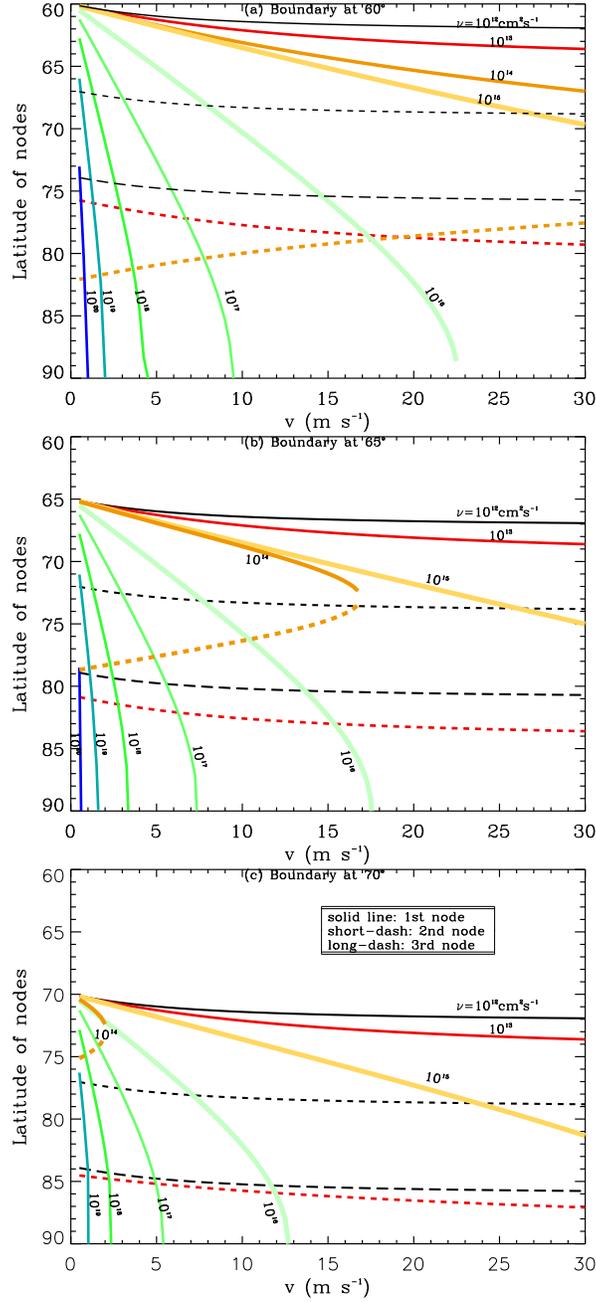}
\caption{Latitudinal positions of streamfunction nodes as a function of 
meridional flow speed at the boundary, for solar type differential rotation
amplitude at the same boundary and a wide range of values of turbulent
viscosity, for boundaries at $60^{\circ}$ (frame a), $65^{\circ}$ (frame b) 
and $70^{\circ}$ (frame c). Solid curves are for the location of the node 
closest to the imposed boundary, short dashed curves the second node, and long 
dashes the third node. The color code is defined by the viscosity values 
shown on each of the solid curves.}
\label{nodeposnu}
\end{figure}
\clearpage

\begin{figure}[hbt]
\epsscale{1.0}
\plotone{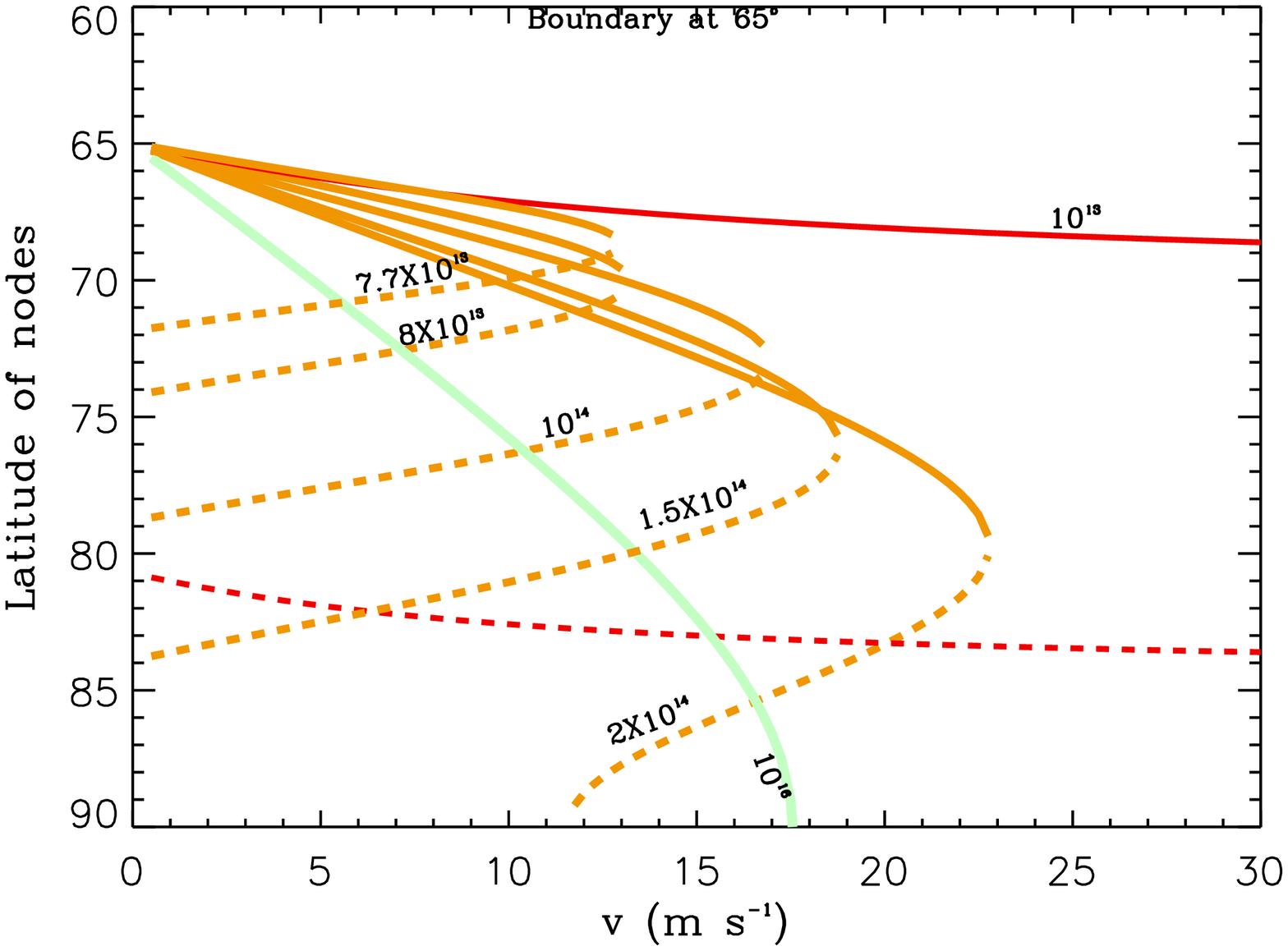}
\caption{Latitudinal positions of streamfunction node for boundary at 
$65^{\circ}$ as a function of meridional flow speed, for turbulent 
viscosities for which an increase of meridional flow speed leads to node 
merger. The color code is defined by the curve labels.}
\label{nodemerge}
\end{figure}

We might conclude from these results that it is virtually impossible to 
generate meridional flow that reaches all the way to the pole for solar like 
viscosity, and that therefore some additional physics would be necessary 
to include to create such flow for solar conditions. But this conclusion 
would be premature, since in frames 5b and c we see evidence of node 'merging'
as the meridional flow is increased. Two nodes merge, eliminating both at
higher meridional flow speeds. This occurs near $\nu=10^{14} {\rm cm}^2
\,{\rm s}^{-1}$ (see orange solid and dashed curves). For the boundary at 
$65^{\circ}$ this merger takes place at a meridional flow speed of about 
$17 {\rm m}\,{\rm s}^{-1}$, and for a boundary at $70^{\circ}$, at about 
$2 {\rm m}\,{\rm s}^{-1}$. Other neighboring boundary placements would 
yield merger flow speeds near these values. We show streamfunctions for 
parameter values near those for which merger occurs in the next section.

Figure 6 gives a more detailed picture of the range of parameter values for
which node merger occurs. Here we display node latitudes for several turbulent
viscosity values in the range $10^{13}-10^{16} {\rm cm}^2\,{\rm s}^{-1}$. 
We see that from $\nu=7.7 \times 10^{13}$ up to $2 \times 10^{14}$ node 
merger occurs, for increased meridional flow speed as the viscosity is 
increased. Above this viscosity range the solution contains only one node, 
similar to the $\nu=10^{16} {\rm cm}^2\,{\rm s}^{-1}$ case shown in green. 
Below $\nu=7.7 \times 10^{13}$, the solutions have two nodes for all 
meridional flow speeds. So for a range of turbulent viscosity of a factor 
of about three, there is node merging in the range of meridional flow speed 
of solar interest.

This phenomenon of node merger occurs for turbulent viscosity values that are 
plausible for the Sun, if somewhat high. Since we know the meridional flow 
speed is observed to vary with time by up to $50\%$, we can easily imagine a 
scenario in which a rise in this speed caused two nodes to merge, allowing 
the primary poleward flow to reach all the way to the pole, as it did for 
most of cycle 23. Thus it may be possible to explain the difference in latitude
to which the primary poleward flow on the Sun reaches using only the physics
we have included here. Since we have made many approximations to get to
the solutions we have found, we should regard the node-mergers shown in
Figure 6 as an example of what is possible, not definitive proof of
an explanation. We anticipate that similar phenomena would occur in the
much more realistic case of a spherical polar cap with a large density
increase downward through the convection zone. That will be explored in a
future paper. A key question to answer will be whether mode-merging in the
spherical case with radial density and viscosity variations occurs for
viscosity values that are more plausible for the Sun.

Figures 5 and 6 do not fully capture the detailed patterns of node location
as a function of the turbulent viscosity. To see these patterns more
clearly, we show in Figure 7 the positions of the first two nodes poleward 
of the boundary as a function of the turbulent viscosity, for a meridional 
flow speed of $10 {\rm m}\,{\rm s}^{-1}$ for boundary placement at 
$60^{\circ}$, $65^{\circ}$ and $70^{\circ}$. In effect, these plots depict 
vertical cuts from Figures 5 and 6 taken at $v=10 {\rm m}\,{\rm s}^{-1}$.

What we see is a rather complex pattern of node locations, divided roughly into
three parts in turbulent viscosity: one pattern with two nodes for 
$10^{11}\leq \nu \leq 2 \times 10^{13} {\rm cm}^2\,{\rm s}^{-1}$, a complex 
transitional pattern in the range $2 \times 10^{13}\leq \nu \leq 3 \times 
10^{14} {\rm cm}^2\,{\rm s}^{-1}$ (the zero node part of which is marked with 
the vertical yellow band), and a pattern with one node or no nodes for  
$\nu \geq 3 \times 10^{14} {\rm cm}^2\,{\rm s}^{-1}$. The transitional 
pattern domain is where node mergers occur, causing the node location and 
number to be quite sensitive to the turbulent viscosity value, as contained 
in the parameter $\epsilon$. We can make sense out of these pattern domains 
by reference to equation (34) for the complex wavenumber $k_n$. If we bring 
the quantity $n^2\pi^2$ outside the square root we get
$$k_n=\pm n\pi (1+(-1)^{1/3}\epsilon_n^{2/3}/n^2\pi^2)^{1/2}.\quad\eqno(47)$$
If we put in solar numbers we find that the expression that includes $\epsilon$
inside the square root is of order one when $\nu$ is in the transition
domain we defined in the neighborhood of $10^{14} {\rm cm}^2\,{\rm s}^{-1}$. 
In physical terms in this range there is a near-equal competition between 
Coriolis forces and turbulent viscous forces. In mathematical terms, in this 
range the phase, or ratio of real to imaginary parts, of several of the roots 
for $k_n$ can change significantly for a small change in $\nu$, leading to 
significantly different node positions and number for neighboring $\nu$ values. 

\begin{figure}[hbt]
\epsscale{1.0}
\plotone{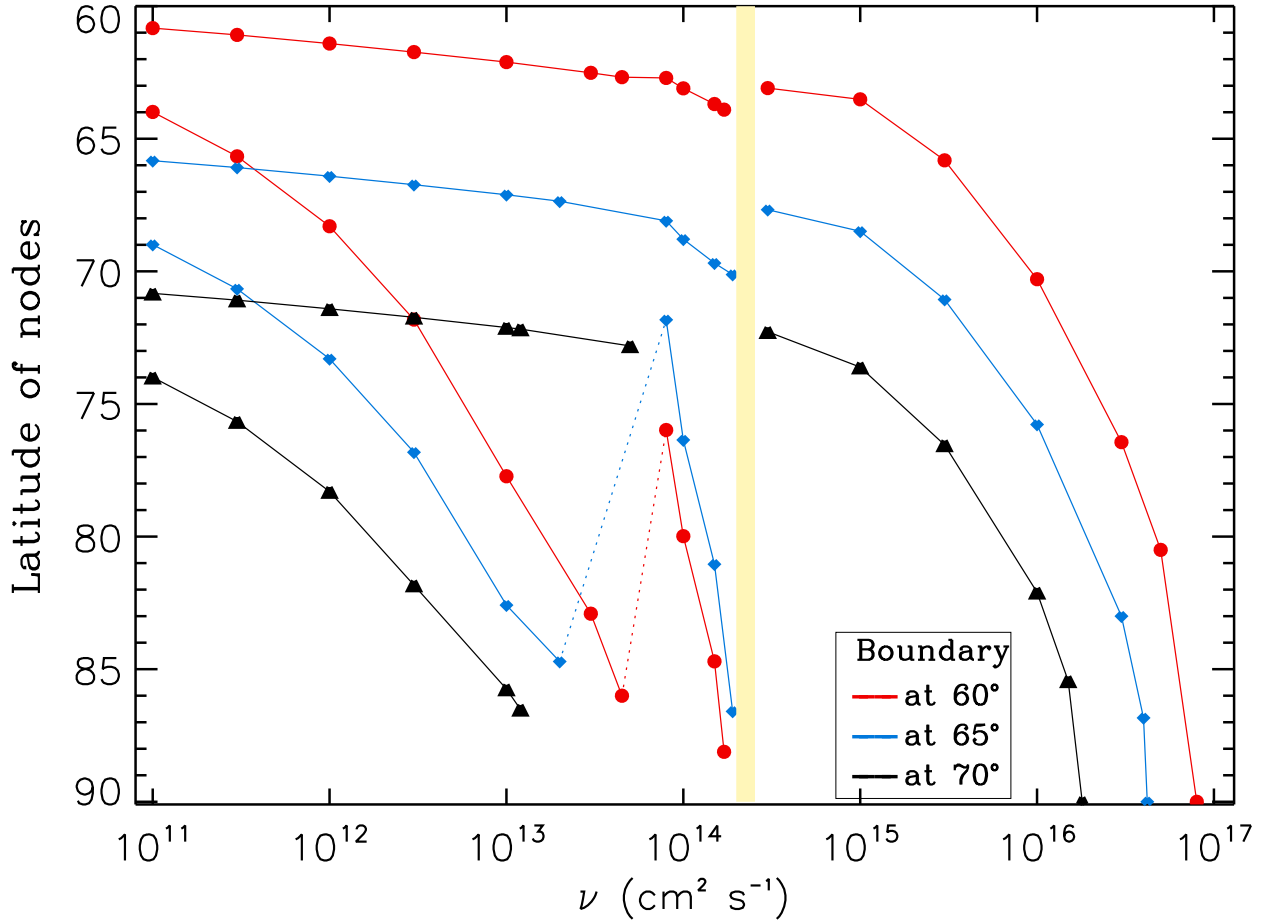}
\caption{Latitudinal positions of streamfunction nodes as a function of 
turbulent viscosity, for a meridional flow speed of $10 {\rm m}\,{\rm s}^{-1}$ 
at the boundary, for solar type differential rotation amplitude at the same 
boundary and three different latitude placements of the boundary. The color 
code is defined in the figure.}
\label{nod1posnu}
\end{figure}

By contrast, when $\nu$ is smaller, the phase of $k_n$ is relatively stable 
and the real and imaginary parts are comparable, leading to a stable pattern
of node number and location. In this situation, it is possible to have two
or even more nodes. As $\nu$ is increased, all nodes migrate to higher
latitudes, because the viscous force is increasingly opposing the Coriolis
force. The position of the second node moves more rapidly to higher latitudes 
with increasing $\nu$ because the amplitude of the polewardmost cell is 
so much weaker than those at lower latitudes.

Finally, when $\nu$ is greater than in the transition domain, the real
part of $k_n$ becomes dominant over the imaginary part, largely eliminating
the oscillatory component of the solutions. The resulting exponential 
functions for the streamfunction are much harder to combine into solutions
with nodes, so there remain only one or zero nodes. These nodes also
migrate toward the poles with increasing $\nu$. In this domain the viscous
forces are totally dominant over the Coriolis forces. As $\epsilon \rightarrow
0$ in equation (47), $k_n \rightarrow \pm n\pi$, leading to purely 
exponential functions as solutions for $\Psi_n$ in this singular limit.
Satisfying the boundary conditions at $x=0,R$ leads to solutions of the form
$\Psi_n(x)=\Psi_n(R)(e^{n\pi x}-e^{-n\pi x})/(e^{n\pi R}-e^{-n\pi R})$. 
By inspection we see that this solution has no node, consistent with the 
approach of the curves tracing position of the first node to $90^{\circ}$
on the right hand side of Figure 7. $\epsilon = 0$ is equivalent to rotation
being zero, so clearly any nodes existent in the solutions must be due to
the action of Coriolis forces.

To what degree does the presence and location of the artificial low latitude
boundary of the determine the latitude of the first node , especially when
it occurs very close to this boundary? We can not know for sure, but we can 
point out that the influence of Coriolis forces will be very important. If
we ignore viscosity, than the latitudinal extent of the meridional flow
beyond the boundary should be limited by the so-called 'inertia circle', which
measures the distance over which a meridional flow isw largely turned into the
direction of rotation by Coriolis forces. Its amplitude is given at high 
latitudes approximately by $v/2\Omega$. For a meridional flow speed of 
$10\,{\rm m}\,{\rm s}^{-1}$, this corresponds to a distance of only about 
$2\times 10^3 \,{\rm km}$, 
extremely short compared to the distance from the boundary to the pole. On
the other hand, as we have pointed out earlier, if rotation is absent, there
is no mechanism to create nodes at all, and the meridional flow reaches all
the way to the pole no matter what the viscosity. Thus in this simple model 
viscosity is needed to break this constraint and allow the flow to reach closer
to the pole.

\subsection{Streamfunction typical cases}

Here we illustrate how the streamfunction patterns evolve as various 
parameters are varied. First we show the full range of meridional flow
cell structures as a function of turbulent viscosity. Then we illustrate
the two quite different ways that the number of streamfunction nodes 
changes as velocity or turbulent viscosity is changed.

Examples of the full range of possible meridional flow structures found with 
our model is displayed in Figure 8; the turbulent viscosity increases from
top to bottom, over the range $10^{12}-10^{17} {\rm cm}^2\,{\rm s}^{-1}$. 
The examples shown range from having three nodes in the streamfunction 
(frames a,b) down to zero nodes (frames g,h). This evolution is accomplished 
first by the countercell expanding to ever higher latitudes, followed by 
the primary cell doing the same. In all cases, the amplitude of each cell 
peaks near its low latitude boundary and each cell successively closer to 
the poles is weaker than its neighbor on the low latitude side. The color 
contours are logarithmic to allow one to see the weaker cells better; from 
the right hand column of frames, it is clear that no matter how many nodes 
there are, on a linear velocity scale, at most only the countercell is 
detectable, and it is always substantially smaller than the primary cell 
that is imposed from low latitudes. These results imply that on the Sun, 
for all viscosity values, it should be possible to detect the countercell, 
but perhaps none of the smaller cells, if any, occurring poleward of it.

\clearpage
\begin{figure}[hbt]
\epsscale{0.8}
\plotone{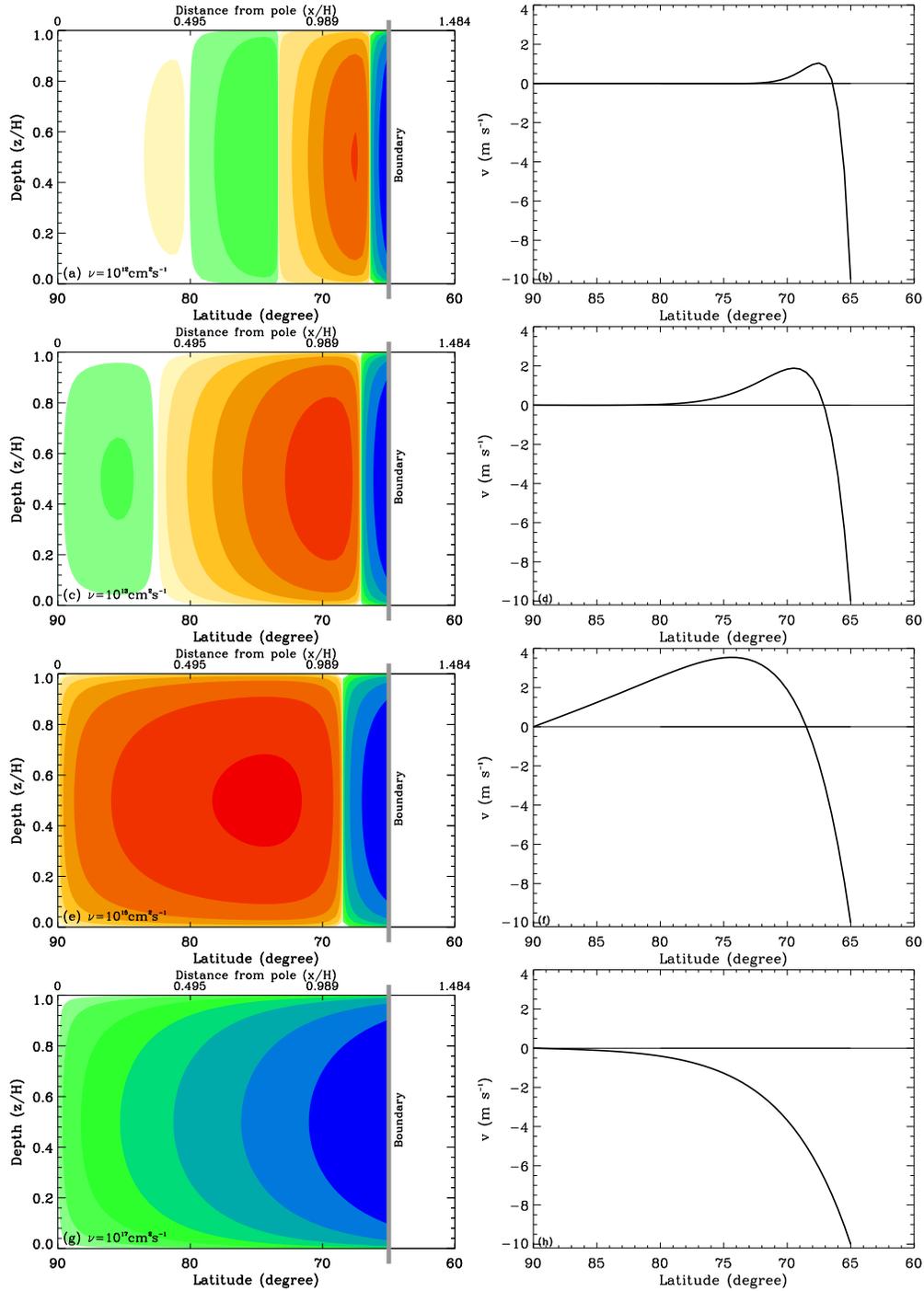}
\caption{Streamfunction contours and line drawings for $v=10 {\rm m}\,
{\rm s}^{-1}$ and boundary at $65^{\circ}$ for increasing values of $\nu$. 
Frames a,b: $10^{12} {\rm cm}^2\,{\rm s}^{-1}$;
frames c,d: $10^{13}$; frames e,f: $10^{15}$; frame g,h: $10^{17}$ that show 
wide range of typical streamfunction patterns. Color code same as Figure 8.} 
\label{typstream}
\end{figure}
\clearpage

\begin{figure}[hbt]
\epsscale{0.39}
\plotone{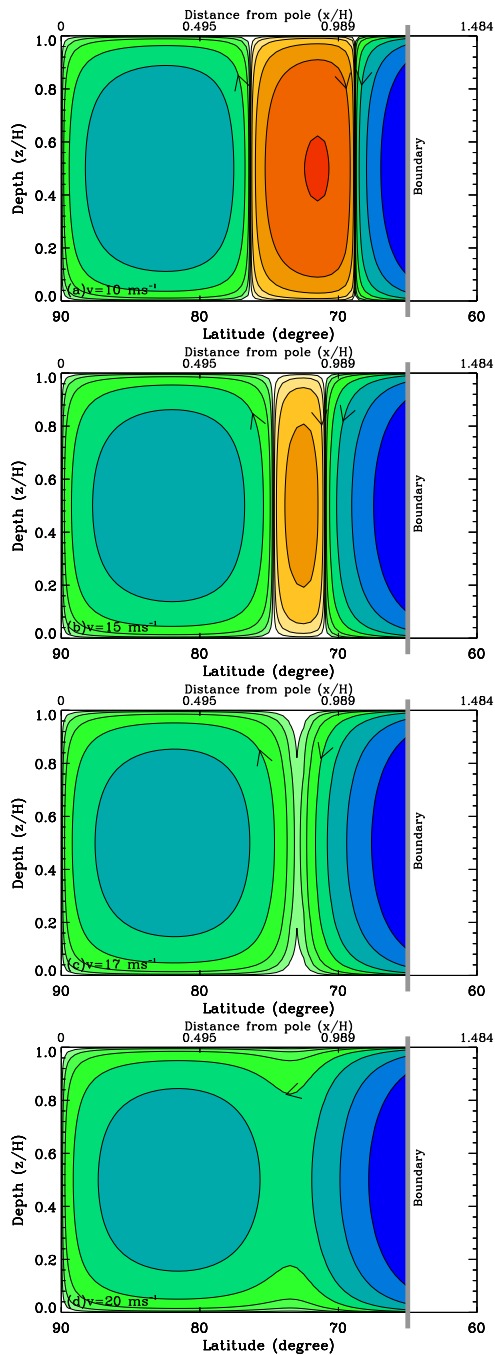}
\caption{Streamfunction contours (logarithmic; red-yellow areas clockwise flow,
green-blue areas counterclockwise flow; arrows indicate the direction of flow)
for a sequence of increasing meridional flow speeds at the boundary at 
$65^{\circ}$, with turbulent viscosity $\nu=10^{14} {\rm cm}^2\,{\rm s}^{-1}$, 
which show the topological changes in the flow patterns as two nodes merge 
and cancel each other out. In frame a, $v=12 {\rm m}\,{\rm s}^{-1}$; frame b, 
$15 {\rm m}\,{\rm s}^{-1}$; frame c, $17 {\rm m}\,{\rm s}^{-1}$; frame d, 
$20 {\rm m}\,{\rm s}^{-1}$.}
\label{nodmergesf}
\end{figure}
\clearpage

\begin{figure}[hbt]
\epsscale{0.45}
\plotone{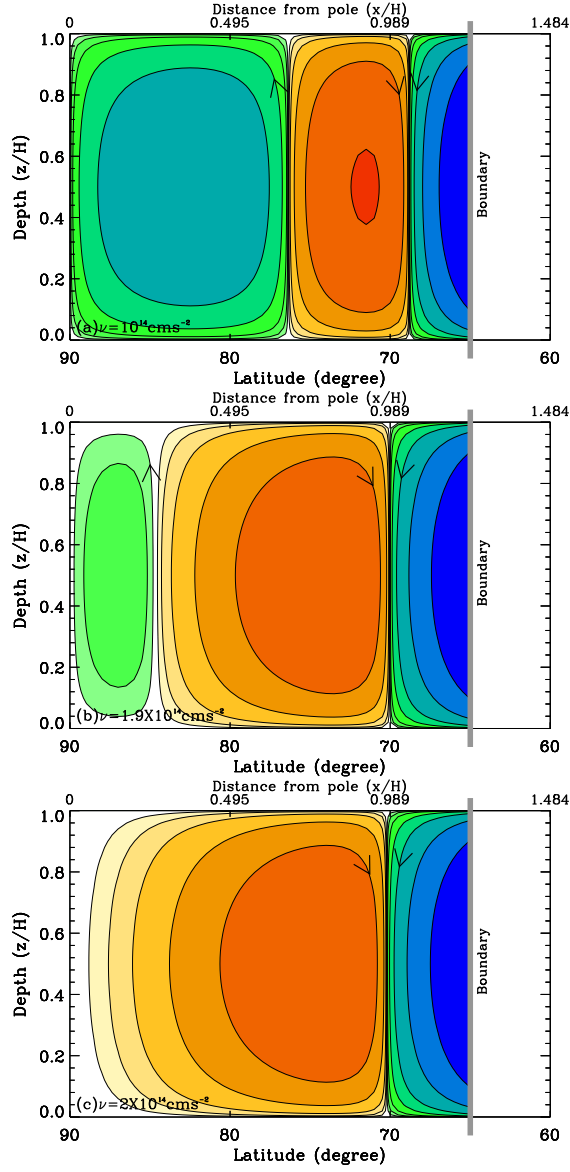}
\caption{Streamfunction contours for $v=10 {\rm m}\,{\rm s}^{-1}$ and boundary 
at $65^{\circ}$ for increasing values of $\nu=10^{14} {\rm cm}^2\,{\rm s}^{-1}$ 
(frame a); $1.8 \times 10^{14}$ (frame b); $2 \times 10^{14}$ (frame c) that 
show evolution of a two node streamfunction pattern into one node as the 
turbulent viscosity is increased, starting from the same case as in frame a 
of Figure 9. Color code same as in Figure 8.}
\label{nodmigr}
\end{figure}
\clearpage

Node merger is displayed in Figure 9. In Figure 9 we can see clearly that 
as the meridional flow speed is increased, node merger, first illustrated in 
Figures 5, 6 and 7, is accomplished topologically by the two counterclockwise 
cells increasing their latitudinal dimension toward each other, until the 
clockwise counter-cell is completely squeezed out and the two counter-clockwise
cells in effect merge to create a single cell that reaches all the way to the 
poles.

By contrast, a very different evolution of the pattern occurs if the boundary
flow speed is fixed and the turbulent viscosity is increased. This is shown
in Figure 10. Here we see that with increasing $\nu$ by just a factor of two, 
starting from the same case as shown in Figure 9a, the larger amplitude 
primary cell and particularly the countercell migrate toward the poles, 
squeezing the second counterclockwise cell out of existence there. The result 
is to retain one node, at a relatively low latitude within the polar cap.
rather than jumping from two nodes to zero. In this example the turbulent
viscosity must be raised two orders of magnitude to eliminate the last node.

We can understand the physics behind this feature if we start with the
definition of viscosity. The viscous coefficient is defined by the ratio of 
shear force per unit area and the gradient of velocity perpendicular to 
the direction of shear. It is inversely proportional to the velocity 
gradient. Therefore with the increase in viscous coefficient the velocity 
gradient should decrease, which means velocities will be more correlated 
over longer distance. Thus each cell grows bigger in the latitude direction
and pushes the next cell, eventually making the polwardmost cell to vanish.

\subsection{Differential rotation}

In Figure 11 we display solutions for the differential rotation linear 
velocity that arises from the forcing at the boundary, for the same cases
for which streamfunction node positions were plotted in Figure 4. We have 
not included in these solutions any differential rotation that is independent 
of $z$ that we might wish to include to allow a closer comparison with solar 
observations, since, as we have stated earlier, this $z$ independent 
differential rotation has no effect on the streamfunctions of meridional 
flow. It is simply a linear function of $x$, declining to zero at the left hand boundary of the cartesian channel from whatever value was specified on the 
right boundary (black straight line). As stated earlier, it corresponds to
constant angular velocity in the cylindrical and spherical shell cases. The 
total solution for differential rotation would be the sum of the appropriate
colored curve and the black line.

\clearpage
\begin{figure}[hbt]
\epsscale{1.0}
\plotone{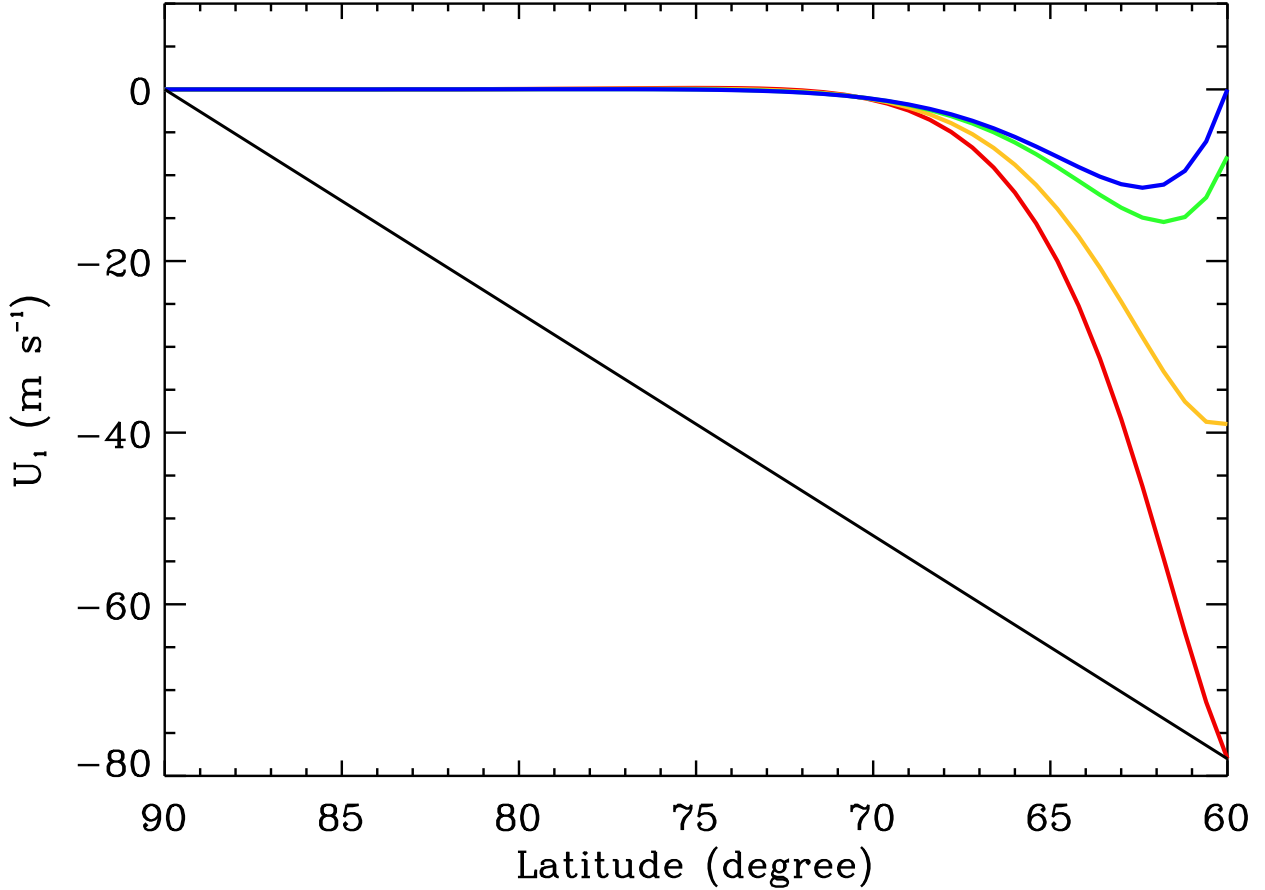}
\caption{Differential rotation linear velocity that occurs in the interior
of the polar cap due to forcing by meridional circulation and differential
rotation at the boundary, for the same cases as shown in Figure 4.  Color
code is the same as in Figure 4. Dimensional linear rotational velocities
are calculated relative to a rotating coordinate system whose rotation is
that of the solar interior below the convection zone. As a consequence
all velocities shown are negative. All linear rotational velocities
are zero at the pole (the left boundary of the cartesian analog). The solid
black straight line running from $-78\,{\rm m}{\rm s}^{-1}$ at $60^{\circ}$ 
to zero at $90^{\circ}$ represents approximately the $z$ independent part of 
the differential rotation linear velocity observed in the Sun at these 
latitudes. In all cases, the differential rotation at $60^{\circ}$ in the 
solution matches that of the boundary there. There is no discontinuity in
the rotation rate there.}
\label{drperturb}
\end{figure}
\clearpage

We already saw from Figure 4 that imposing a $z$-dependent differential
rotation on the boundary has rather little effect on the positions of nodes
in the streamfunction. In Figure 11 we see that the perturbation in 
differential rotation does not extend that far into the polar cap domain. 
When no differential rotation is imposed (blue curve) a drop in linear rotation
is produced that is confined to the first $10^{\circ}$ of the forcing boundary.
This arises due to the Coriolis force from the meridional flow, which itself 
is declining in amplitude with distance polarward from the forcing boundary.
The amplitude of this differential rotation is about that of the observed
torsional oscillations.

For much higher forcing by differential rotation at the boundary, this 
structure is overwhelmed by the simple viscous damping of the rotational flow 
with poleward distance from the boundary. Coriolis forces from the meridional 
flow can not maintain these higher values, and the Coriolis force from the 
imposed differential rotation has only a minor effect on the meridional flow.
As a point of comparison, the observed $z$-independent linear differential
rotation at high latitudes is given by a straight diagonal line (in black) from
the lower right corner of Figure 11, up to the zero point on the left
axis, in accordance with the solutions that contain only differential rotation
discussed in section 3.6.1. The large difference between this amplitude and 
that of the four colored curves shown shows how small the differential rotation
in our $n=1$ solutions is, except very close to the boundary where the forcing
is applied.

\section{Discussions and conclusions}

We have developed a relatively simple hydrodynamical model of the circulation
at high latitudes in the solar convection zone and photosphere that contains
only three forces: pressure gradients, viscous and Coriolis forces. The model
equations are solved in a cartesian 'analog' of a spherical polar
cap that leaves out curvature effects as well as the large density increase 
with depth. This system is assumed to be stress-free at the top and 
bottom, corresponding to the top and bottom of the solar convection zone.
It is forced with meridional flow and differential rotation, guided by
observations, imposed at the low latitude boundary of the cap, placed at 
latitudes between $60^{\circ}$ and $70^{\circ}$ latitude. While the inclusion
of such a boundary is artificial, it is intended to separate the physics
of low and mid-latitudes, responsible for the primary meridional circulation
cell that has poleward flow at the top, from the physics active at high
latitudes, which should be the primary determinant of the circulation found
there.

Our general results are that, as the turbulent viscosity is increased, the
number of nodes decreases. As the meridional flow is increased for a
given turbulent viscosity, the latitude of a node increases; or, in other words,
the node migrates poleward. The first of 
these results is explained by the viscous forces increasingly overpowering 
the deflecting effect of the coriolis force to allow the poleward flow of 
the primary cell to reach a higher latitude. The second is due to the 
increased poleward momentum of the poleward moving particles that allows them
to reach a higher latitude before sinking down to feed the return flow.
Overall, we find that our general results are not particularly sensitive
to the differential rotation imposed at the boundary, provided it is
plausible for the Sun. Unlike for the meridional circulation, changes in
the differential rotation of the Sun at all latitudes are very small
percentages of the mean differential rotation.

Most interestingly, we find that the decrease in the number of nodes as
the turbulent viscosity is increased is not monotonic. In particular, in
the neighborhood of $\nu=10^{14} {\rm cm}^2\,{\rm s}^{-1}$ in this model, 
we find that two nodes merge as the meridional flow is increased, leaving 
no nodes for higher meridional flow at the boundary. This is true even though, 
for still higher viscosity, there remains one node. Two nodes implies the
presence of both a reversed (clockwise) cell and, on its poleward side, a 
second, weak, counterclockwise cell that has a poleward flow near the top, 
just as in the primary cell. With the merger of the two nodes, the reverse 
cell is squeezed out, merging the primary cell with its polar counterpart. 

We speculate that it is this phenomenon that could be responsible for the
primary cell reaching all the way to the poles during much of cycle 23,
whereas there was a node near $65^{\circ}$ in both cycles 22 and the early
stages of cycle 24. At present, observations can not tell us whether there
is a second node near $80^{\circ}$ as our model predicts. Better observations
in the future of velocities at the highest latitudes would obviously be 
valuable for testing this theory and that of the spherical polar cap that
will be developed later. Better information about variations of meridional
flow with depth at high latitudes would also be very useful.

Our solutions are for steady flow, but we know that the meridional flow
at high latitudes on the Sun changes with time. Given our results, we can
expect that a time dependent theory could determine whether changes in
meridional flow at the boundary would lead to changes in meridional flow
at higher latitudes that agree with observed changes in flow at the highest 
latitudes. If the Sun merges two nodes with a meridional flow increase,
does the model do the same? Does the model predict a merger as a result
of an increase in meridional flow that is not observed on the Sun? Questions
such as these can only be answered with a time dependent model.

Our results suggest that it may not be necessary to invoke any additional 
physics to explain changes in the meridional flow cells with time in high
latitudes. That does not mean, however, that the presence or importance of
such physics can be ruled out at this time.

There are several additional effects that should be included in the high
latitude meridional circulation model to make it more realistic. Even keeping
the same physics, results from this model using spherical geometry could
change significantly. With spherical geometry the meridians converge to the
pole, making it harder for as much mass flux to reach the pole as does in the 
straight channel. In addition, the spherical problem does not separate easily
in radius and colatitude due to the Coriolis forces. This has the effect of
linking different latitudes and different depths of the flow in ways not 
present in the cartesian analog. All of these spherical effects should 
influence the structure of the flow, including the location of nodes in the 
streamfunction, and how many nodes there are for a given turbulent viscosity.

Within spherical geometry, the flow patterns will change substantially when
the density increase with depth through the convection zone is included;
the patterns of mass flux, or $\rho v$, could look somewhat like those
without the density variation, but the velocity itself should decline
substantially with depth. This effect could also change the location and number
of nodes present. In this version of the model, effects of the variation of
the turbulent viscosity with depth should also be studie; we have already seen
that the number and latitude of nodes in the meridional flow is sensitive to
the turbulent viscosity used. Allowing this quantity to vary with radius in
the spherical case could change the results substantially, as could allowing
for the density increase with depth.

In its present form, the model is for axisymmetric motions, and, beyond the
turbulent diffusion, our model contains no effect of organized global
scale Reynolds stresses. Yet these stresses are surely important in driving
the global differential rotation as well as playing a role in the 
maintenance of the primary meridional cell. We do not know to what latitude
these Reynolds stresses reach, but they could be active within the polar cap.
Their possible effects on the polar meridional flows should also be considered.

Our model currently also does not include explicitly any thermodynamics. 
Allowing for departures of the temperature from the adiabatic gradient
could be important, particularly at the top and the bottom of the convection
zone. Finally, our model is hydrodynamic, so no effects of magnetic fields
are included. But the Sun is a dynamo which generates fields, some quite
strong, throughout the convection zone, so the effects of these should also
be taken into account. One possible effect is that different amplitudes
of polar fields might influence the amplitude of meridional flow in
high latitudes, as well as the locations of its nodes.

\acknowledgements

We thank Nick Featherstone for reviewing the entire manuscript and for
his helpful comments. We extend our thanks to an anonymous reviewer
for a thorough review and for constructive comments, which have helped
us improve the paper significantly. This work is partially supported 
by NASA's Living With a Star program through the grant NNX08AQ34G. 
The National Center for the Atmospheric Research is sponsored by the 
National Science Foundation.

\end{document}